\documentclass[preprint,12pt]{elsarticle}

\usepackage{graphicx}

\usepackage{amssymb}
\usepackage{amsmath}

\usepackage{xcolor}
\def\drawline#1#2{\raise 2.5pt\vbox{\hrule width #1pt height #2pt}}
\def\spacce#1{\hskip #1pt}
\def\solid{\drawline{24}{.5}\nobreak\ }
\def\bdash{\hbox{\drawline{4}{.5}\spacce{2}}}
\def\dashed{\bdash\bdash\bdash\bdash\nobreak\ }
\def\bdot{\hbox{\drawline{1}{.5}\spacce{2}}}
\def\dotted{\hbox{\leaders\bdot\hskip 24pt}\nobreak\ }
\def\chndash{\hbox {\drawline{8.5}{.5}\spacce{2}\drawline{3}{.5}\spacce{2}\drawline{8.5}{.5}}\nobreak\ }
\def\chndot{\hbox {\drawline{9.5}{.5}\spacce{2}\drawline{1}{.5}\spacce{2}\drawline{9.5}{.5}}\nobreak\ }

\def\chndashdash{\hbox {\drawline{8}{.5}\spacce{2}\drawline{3}{.5}\spacce{2}\drawline{3}{.5}\spacce{2}\drawline{8}{.5}}\nobreak\ }

\journal{Journal of Computational Physics}

\begin{document}

\begin{frontmatter}



\title{A finite--difference scheme for three--dimensional incompressible
flows in spherical coordinates}


\author{L. Santelli$^1$, P. Orlandi$^2$ \& R. Verzicco$^{1,3,4}$}

\address{$^1$ Gran Sasso Science Institute, L'Aquila, Italy\\
         $^2$ Sapienza Universit\`a di Roma, Roma, Italy\\
         $^3$ Universit\`a di Roma `Tor Vergata', Roma, Italy\\
         $^4$ PoF, University of Twente, Enschede, The Netherlands}

\begin{abstract}
In this study we have developed a flexible and efficient numerical scheme for the
simulation of three--dimensional incompressible flows in spherical coordinates. 
The main idea, inspired by a similar strategy \textcolor{black}{as} \cite{Verzi96} for cylindrical
coordinates, consists of a change of variables combined with a discretization
on a staggered mesh \textcolor{black}{and the special treatment of few discrete
terms} that remove the singularities of the Navier--Stokes equations
at the sphere centre and along the polar axis. 
\textcolor{black}{
This new method alleviates also the time step restrictions 
introduced by the discretization around the polar axis while 
the sphere centre still yields strong limitations, 
although only in very unfavourable flow configurations.
}

The scheme is second--order accurate in space and is verified and validated
by computing numerical examples that are compared with similar results produced by
other codes or available from the literature.

The method can cope with flows evolving in the whole sphere, in a spherical shell and
in a sector without any change and, thanks to the flexibility of finite--differences,
it can employ generic mesh stretching (in two of the three directions) and complex 
boundary conditions.

\end{abstract}


\end{frontmatter}


Thanks to the growing availability of computational power also the 
complexity of flows tackled by numerical simulations is increasing.
Among many, one of the challenges of a computation is the
mathematical description of non trivial domains and the
flows developing within spherical geometries belong to this category.

\begin{figure}[htb!]
\begin{center}
\includegraphics[width=2.25in]{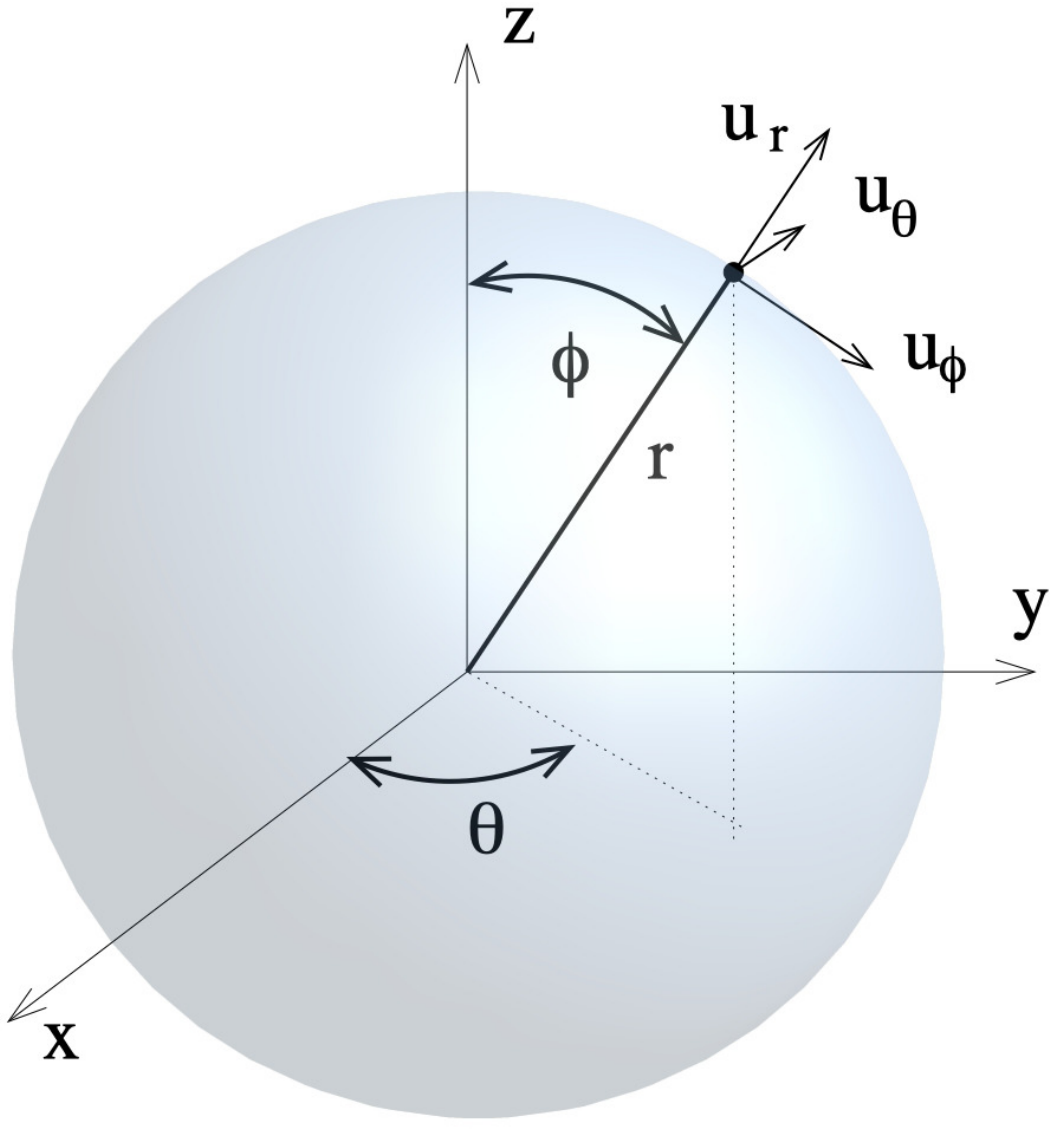}
\end{center}
\caption{Sketch of the system and coordinate definition.}
\label{fig1n}
\end{figure}

Indeed, the mapping of spherical to Cartesian coordinates reads
$
x = r \cos \theta \sin \phi, \quad
y = r \sin \theta \sin \phi, \quad
z = r \cos \phi,
$
(Fig. \ref{fig1n}) and it is not single valued at the centre ($r=0$) and
on the polar axis ($\phi=0$, $\pi$) therefore, even if
the sphere is among the simplest shapes, its natural
coordinate system contains mathematical singularities that are reflected
also in the governing equations for fluid flows (see next section).

\textcolor{black}{
Spherical domains have been
 traditionally used in geophysics \cite{Rolf11}, 
oceanography \cite{Const18}, meteorology \cite{Rand02},
astrophysics \cite{Akse18}, \cite{Glatz84} and
 magnetohydrodynamics \cite{Kiss18} although they are gaining
popularity also for  
industrial \cite{Osch18} and fundamental problems \cite{Busse75}, \cite{Kida94}, \cite{Seyb18}.
}

In fact, an important advantage of the spherical coordinates is that 
their highest degree of symmetry imposes no preferred orientation and this is a desired
property when the system evolves in an unbounded space \cite{Boden07}.
That the mesh topology could interfere with the flow dynamics was shown by
 \cite{Verzi96} who computed the evolution of an azimuthally unstable 
toroidal vortex ring on a Cartesian square mesh and obtained preferentially the 
$n=4$ wavenumber even if the linear stability analysis predicted the $n=5$ mode. 
On the other hand, the same
phenomenon, simulated using a polar cylindrical mesh \cite{Orla93}, correctly 
showed the emergence of fivefold symmetric structures as expected from the theory \cite{Shar94}.

Since the early stability analyses of thermal convection in spherical shells \cite{Busse75},
also in later studies \cite{Glatz84}, \cite{Forn95}, \cite{Kitau97}, \cite{Tilgn99}
 the variables were expressed by spherical harmonics for the longitude $\theta$ and
colatitude $\phi$ while Chebyshev polynomials were mostly 
used for the radial direction $r$ (Fig. \ref{fig1n}).
This expansion allowed the development of accurate pseudo--spectral methods that
avoided the equation singularity at the polar axis.

All these studies, however, did not cope with the singularity at the sphere centre ($r=0$) 
since the equations were solved only in the gap between two concentric spheres.

On the other hand, \cite{Kagey00} developed a spectral method that could discretize
the equations up to the sphere centre by using only the even order Chebyshev polynomials 
for the radial direction that avoided the unnecessary mesh refinement near the origin.
Also \cite{Kida08} simulated the whole sphere although they used Jacoby polynomials that,
in addition to the previous property, satisfy also the regularity conditions at the poles.
Using this numerical scheme \cite{Kida11} were able to simulate magnetohydrodynamic
processes in a precessing sphere.
\cite{Auter09} developed a Navier--Stokes spectral solver in a sphere employing a
latitude dependent number of modes that avoided the difficulties with the sphere centre
and polar axis.

Spectral methods are generally preferred to other approaches since, although the 
overall accuracy depends also on the dealiasing schemes, they yield smaller numerical
errors for a given number of nodes (modes).
However,
as shown by \cite{Moin16}, finite--difference approximations, if properly implemented,
 become very competitive cost--wise with \textcolor{black}{respect to} spectral methods even considering that the former 
 require more computational nodes to achieve the same precision.
In addition, if the simulation has to
cope with complex boundary conditions, variable fluid properties or 
generic node
distributions then finite--differences are the best option.

Despite these advantages,
the literature on finite--difference methods for the solution of the Navier--Stokes
equations in spherical coordinates is scarce and, to the knowledge of the authors
only few studies are available. 
\cite{Willia73} presented a method for solving the equations in a spherical
shell by finite--difference approximations on a staggered mesh in all three directions;
the variables were modified at the poles by semi--analytic corrections to avoid
stability issues.
\cite{Gilma75} solved the linear Boussinesq convection
in a spherical shell
using Fourier modes along the latitude $\theta$ and finite--differences in 
$r$ and $\phi$. The model was completed with  the nonlinear convective terms
in \cite{Gilma77} and, in both cases, in order to avoid the stability limitations
at the polar axis, a low--pass filtering of the solution was applied
around the poles. Apparently, this smoothing strategy was widely adopted in the
atmospheric community and \cite{Cullen83} described it in a systematic way.

\cite{Kagey93} relied on finite--difference approximations in all three directions
for the simulation of convection in a rotating spherical shell and, also in this case,
a low--pass filter near the poles was employed.

\cite{Aube08} used the code PARODY to simulate convection--driven numerical dynamos
in a spherical shell; in this case 
second--order finite differences were used only for the radial directions
while spherical harmonics were adopted for the
lateral ones.
The scheme was very similar to that of \cite{Glatz84} although the new radial
discretization made the code suitable for parallel computation on distributed memory 
clusters.

\textcolor{black}{In the paper \cite{Sha98}, second--order finite--differences 
are employed to solve the Navier-Stokes equations in 
spherical coordinates relying only on the semi--conservative form of the equations and on the discretization on a staggered-- mesh to remove the singularities.
However, only flows within two spherical shells are considered, thus never 
copying with the singularity at r=0. In addition, all their examples had 
vanishing
meridional velocity at the poles therefore the singularities of the equations 
were not really tackled.}

\textcolor{black}{
\cite{Mohs00} proposed a clever procedure to avoid the polar singularity 
(both, cylindrical and spherical) by extending the radial coordinate
to negative values and discretizing the domain so that no nodes are located
at $r=0$. They consider the compressible Navier--Stokes equations and use a 
co--located discretization with 
high order finite--differences for the radial direction and spectral methods
in the remaining ones. The application of the same method to a fully 
staggered discretization for the incompressible Navier--Stokes equations 
is not obvious and it would presumably require substantial changes.
}

The study by \cite{Boers98} uses a finite--volume method to solve the Navier--Stokes 
equations in spherical coordinates.  In this case, however, neither the singularity at
the centre nor at the polar axis is encountered since the domain of interest is restricted
to a spherical sector in which the self--similar region of a round jet is computed.

In this paper we present a novel numerical method for the solution of the incompressible
Navier--Stokes equations in spherical coordinates. It is 
based on second--order finite--difference
approximations on a staggered mesh that, combined with a change of variables
\textcolor{black}{and a special treatment of some discretized terms}, eliminates
the singularities at the polar axis and at the sphere centre, simultaneously. 
The same method can be applied both, to flows developing in a spherical shell and in
the whole sphere up to $r=0$ without any change in the numerical procedure.

\textcolor{black}{
The time step restrictions 
introduced by the discretization around the polar axis and
sphere centre are attenuated for the former region while
the latter still gives strong time step limitations, 
although only in very unfavourable cases with the largest flow
velocity occuring at $r=0$.
}

We show that the method maintains the second--order accuracy and yields 
free--divergence velocity fields to machine precision even when using computational
meshes that are unnecessarily refined at the polar axis and sphere centre.

The method is verified and validated by computing numerical examples that stress the
treatment of the equations at the singular points and by comparing the results with
analogous computations available from the literature.

Finally, since the proposed method takes after the scheme of \cite{Verzi96} it shares
the same variable arrangement and memory layout, therefore it is efficiently and 
massively parallelized as done in \cite{Steve10}.

The paper is organized as follows: in the next section we present the equations
and the change of variables adopted to remove the singularity.
In section \ref{sec:vd} we discuss the discretization of the variables and the 
technicalities needed for some terms while in section \ref{sec:nm} we briefly
describe the numerical method. 
In section \ref{sec:re} a number of numerical examples is shown and discussed
to assess the reliability and efficiency of the method. Finally in section
\ref{sec:co} the closing remarks and perspectives for future work are given.


\section{The equations}
\label{sec:eq}

The continuity and momentum equations for an incompressible and viscous flow,
in non dimensional form and in spherical coordinates, read \cite{Batch67}:
\begin{equation}
\frac{1}{r^2}\frac{\partial r^2 u_r}{\partial r} +
\frac{1}{r \sin \phi}\frac{\partial u_\theta }{\partial \theta}+
\frac{1}{r \sin \phi}\frac{\partial \sin \phi u_\phi }{\partial \phi} = 0,
\label{eq:con}
\end{equation}
$$
\frac{\partial u_r}{\partial t} +{\bf u}\cdot \nabla u_r -\frac{u_\phi^2}{r}
-\frac{u_\theta^2}{r} = -\frac{\partial p}{\partial r} + f_r+
$$
$$
\frac{1}{Re} \left (
\nabla^2 u_r -\frac{2 u_r}{r^2} -\frac{2}{r^2\sin \phi}\frac{\partial u_\phi \sin \phi}{\partial \phi}
-\frac{2}{r^2\sin^2 \phi}\frac{\partial u_\theta}{\partial \theta} \right ),
$$
$$
\frac{\partial u_\theta}{\partial t} +{\bf u}\cdot \nabla u_\theta +\frac{u_\theta u_r}{r}
+\frac{u_\theta u_\phi}{r\tan \phi } = -\frac{1}{r \sin \phi} \frac{\partial p}{\partial \theta} +f_\theta
$$
$$
+ \frac{1}{Re} \left (
\nabla^2 u_\theta +\frac{2 }{r^2\sin^2 \phi}\frac{\partial u_r}{\partial \theta} +\frac{2\cos \phi}{r^2\sin^2 \phi}\frac{\partial u_\phi}{\partial \theta} - \frac{u_\theta}{r^2 \sin^2 \phi}
 \right ),
$$
$$
\frac{\partial u_\phi}{\partial t} +{\bf u}\cdot \nabla u_\phi +\frac{u_\phi u_r}{r}
-\frac{ u_\theta^2}{r\tan \phi } = -\frac{1}{r} \frac{\partial p}{\partial \phi}+f_\phi
$$
\begin{equation}
+ \frac{1}{Re} \left (
\nabla^2 u_\phi +\frac{2 }{r^2}\frac{\partial u_r}{\partial \phi} -\frac{u_\phi}{r^2\sin^2 \phi} - \frac{2 \cos \phi}{r^2 \sin^2 \phi}\frac{\partial u_\theta}{\partial \theta}
 \right ),
\label{eq:ns}
\end{equation}

where $u_\theta$, $u_r$ and $u_\phi$ are the velocity components in the longitudinal,
radial and colatitude directions, respectively, $p$ the pressure and 
$f_\theta$, $f_r$ and $f_\phi$ forcings that could be used for volume forcings
such as Coriolis accelerations.
$Re = UL/\nu$ is the Reynolds number defined by appropriate velocity $U$ and length $L$
scales and $\nu$ is the kinematic viscosity of the fluid.

In the above equations the following relations are used:
$$
{\bf u}\cdot \nabla q \equiv u_r \frac{\partial q}{\partial r}
+\frac{ u_\theta}{r\sin \phi } \frac{\partial q}{\partial \theta} 
+\frac{ u_\phi}{r } \frac{\partial q}{\partial \phi} 
$$
and
$$
\nabla^2 q \equiv \frac{1 }{r^2}\frac{\partial }{\partial r} r^2\frac{\partial q}{\partial r} 
+ \frac{1}{r^2 \sin \phi}\frac{\partial }{\partial \phi}\sin \phi \frac{\partial q}{\partial \phi}
+ \frac{1}{r^2 \sin^2 \phi}\frac{\partial^2 q }{\partial \theta^2}.
$$

As anticipated in the Introduction, many terms of Equations (\ref{eq:con}--\ref{eq:ns})
become singular at the origin ($r=0$) and at the North and South poles, respectively, 
$\phi = 0$ and $\phi = \pi$ (hereinafter referred to as `polar axis') and this is
not due to the physics described by the equations but to the spherical coordinate 
transformation that is not single valued at those points.
Another drawback is that the coordinates $r=0$ and $\phi = 0$, $\pi$
do not coincide with physical boundaries, such as a solid wall or a slip surface,
therefore boundary conditions for the unknowns can not be easily computed there.
The same argument does not apply to the longitudes
$\theta = 0$, $2\pi$ that, being the same physical point, can benefit from periodic 
boundary conditions that do not need explicit values for the unknowns.

Following \cite{Verzi96} and motivated by the above arguments, we introduce a new
set of unknowns
${\bf q} = (q_\theta, q_r, q_\phi ) = (u_\theta, u_r r^2,u_\phi \sin \phi)$ 
that, according to Equation (\ref{eq:con}), can also be thought of as volume
fluxes.
An immediate advantage is that 
these variables yield $q_r(\theta,0,\phi) = q_\phi(\theta,r,0) = q_\phi(\theta,r,\pi) \equiv 0$
therefore transforming the difficult singular points for ${\bf u}$ in trivial boundary conditions
for ${\bf q}$. 

By proper manipulation of Equations (\ref{eq:con}--\ref{eq:ns}) they can be easily 
rewritten 
in terms of the new variables ${\bf q}$:
\begin{equation}
\sin \phi \frac{\partial q_r}{\partial r} +
r \frac{\partial q_\theta }{\partial \theta}+
r \frac{\partial q_\phi }{\partial \phi} = 0,
\label{eq:conq}
\end{equation}

\vskip 0.25 cm

$$
\frac{\partial q_\theta}{\partial t} 
+\frac{1}{r^2}\frac{\partial q_r q_\theta }{\partial r} 
+\frac{ 1}{r\sin \phi } \frac{\partial q_\theta^2 }{\partial \theta} 
+\frac{ 1}{r \sin \phi} \frac{\partial q_\theta q_\phi }{\partial \phi}
+\frac{q_\theta q_r}{r^3}
+\frac{q_\theta q_\phi}{r\tan \phi \sin \phi} 
= -\frac{1}{r \sin \phi} \frac{\partial p}{\partial \theta} +f_\theta
$$
$$
+ \frac{1}{Re} \left (
\frac{1}{r^2}\frac{\partial }{\partial r} r^2\frac{\partial q_\theta}{\partial r} 
+ \frac{1}{r^2 \sin \phi}\frac{\partial }{\partial \phi}\sin \phi \frac{\partial q_\theta}{\partial \phi}
+ \frac{1}{r^2 \sin^2 \phi}\frac{\partial^2 q_\theta }{\partial \theta^2}
\right .
$$
\begin{equation}
\left .
+\frac{2 }{r^4\sin \phi}\frac{\partial q_r}{\partial \theta} +\frac{2\cos \phi}{r^2\sin^3 \phi}\frac{\partial q_\phi}{\partial \theta} - \frac{q_\theta}{r^2 \sin^2 \phi}
 \right ),
\tag{4a}
\label{eq:qth}
\end{equation}

\vskip 0.25 cm

$$
\frac{\partial q_r}{\partial t} 
+\frac{\partial }{\partial r} \left ( q_r \frac{q_r}{r^2} \right )
+\frac{ 1}{r\sin \phi } \frac{\partial q_r q_\theta }{\partial \theta} 
+\frac{ 1}{r \sin \phi} \frac{\partial q_r q_\phi }{\partial \phi}
-\frac{r q_\phi^2}{\sin^2 \phi} 
-r q_\theta^2 = -r^2 \frac{\partial p}{\partial r} + r^2 f_r
$$
$$
\frac{1}{Re} \left (
\frac{\partial }{\partial r} r^2\frac{\partial q_r/r^2}{\partial r} 
+ \frac{1}{r^2 \sin \phi}\frac{\partial }{\partial \phi}\sin \phi \frac{\partial q_r}{\partial \phi}
+ \frac{1}{r^2 \sin^2 \phi}\frac{\partial^2 q_r }{\partial \theta^2}
\right .
$$
\begin{equation}
\left .
-\frac{2 q_r}{r^2} -\frac{2}{\sin \phi}\frac{\partial q_\phi }{\partial \phi}
-\frac{2}{\sin \phi}\frac{\partial q_\theta}{\partial \theta} \right ),
\tag{4b}
\label{eq:qr}
\end{equation}

\vskip 0.25 cm

$$
\frac{\partial q_\phi}{\partial t} 
+\frac{1}{r^2}\frac{\partial q_r q_\phi }{\partial r} 
+\frac{ 1}{r\sin \phi } \frac{\partial q_\theta q_\phi }{\partial \theta} 
+\frac{ 1}{r } \frac{\partial }{\partial \phi} \left ( q_\phi \frac{q_\phi}{\sin \phi} \right )
+\frac{q_\phi q_r}{r^3} 
-\frac{ q_\theta^2 \cos \phi}{r } = -\frac{\sin \phi}{r} \frac{\partial p}{\partial \phi} 
+ \sin \phi f_\phi
$$
$$
+ \frac{1}{Re} \left (
\frac{1}{r^2}\frac{\partial }{\partial r} r^2\frac{\partial q_\phi}{\partial r} 
+ \frac{1}{r^2 }\frac{\partial }{\partial \phi}\sin \phi \frac{\partial q_\phi/\sin \phi}{\partial \phi}
+ \frac{1}{r^2 \sin^2 \phi}\frac{\partial^2 q_\phi }{\partial \theta^2}
\right .
$$
\begin{equation}
\left .
+\frac{2 \sin \phi}{r^2}\frac{\partial q_r/r^2}{\partial \phi} -\frac{q_\phi}{r^2\sin^2 \phi} - \frac{2 \cos \phi}{r^2 \sin \phi}\frac{\partial q_\theta}{\partial \theta}
 \right ),
\tag{4c}
\label{eq:qph}
\end{equation}

that now can be discretized on a computational grid.

\textcolor{black}{
It is worthwhile noticing that the above equations, once discretized on a 
staggered--mesh, are equivalent to those for the contravariant velocity 
components (multiplied by the cell volume) in general curvilinear coordinates
as shown in \cite{Rose91}.
 However, the equations in general curvilinear coordinates entail several 
metric terms that add to the operation count and to the storage requirement, 
leaving aside the augmented data--transfer across nodes in parallel computing. 
On the other hand, in spherical coordinates the metric terms reduce to a 
one--dimensional vector for the radial direction and a two--dimensional array 
(obtained by the product of two one--dimensional vectors) for the azimuthal 
direction \cite{Batch67} thus largely reducing the mentioned drawbacks. 
}

\section{Variable discretization}
\label{sec:vd}

Equations (\ref{eq:conq}, \ref{eq:qth}--\ref{eq:qph}) are discretized by central second--order accurate 
finite--difference approximations along the same line as \cite{Verzi96}.
Here we describe the technicalities needed for the calculation of
some representative terms.  

\begin{figure}[htb!]
\begin{center}
\includegraphics[width= 0.95 \textwidth]{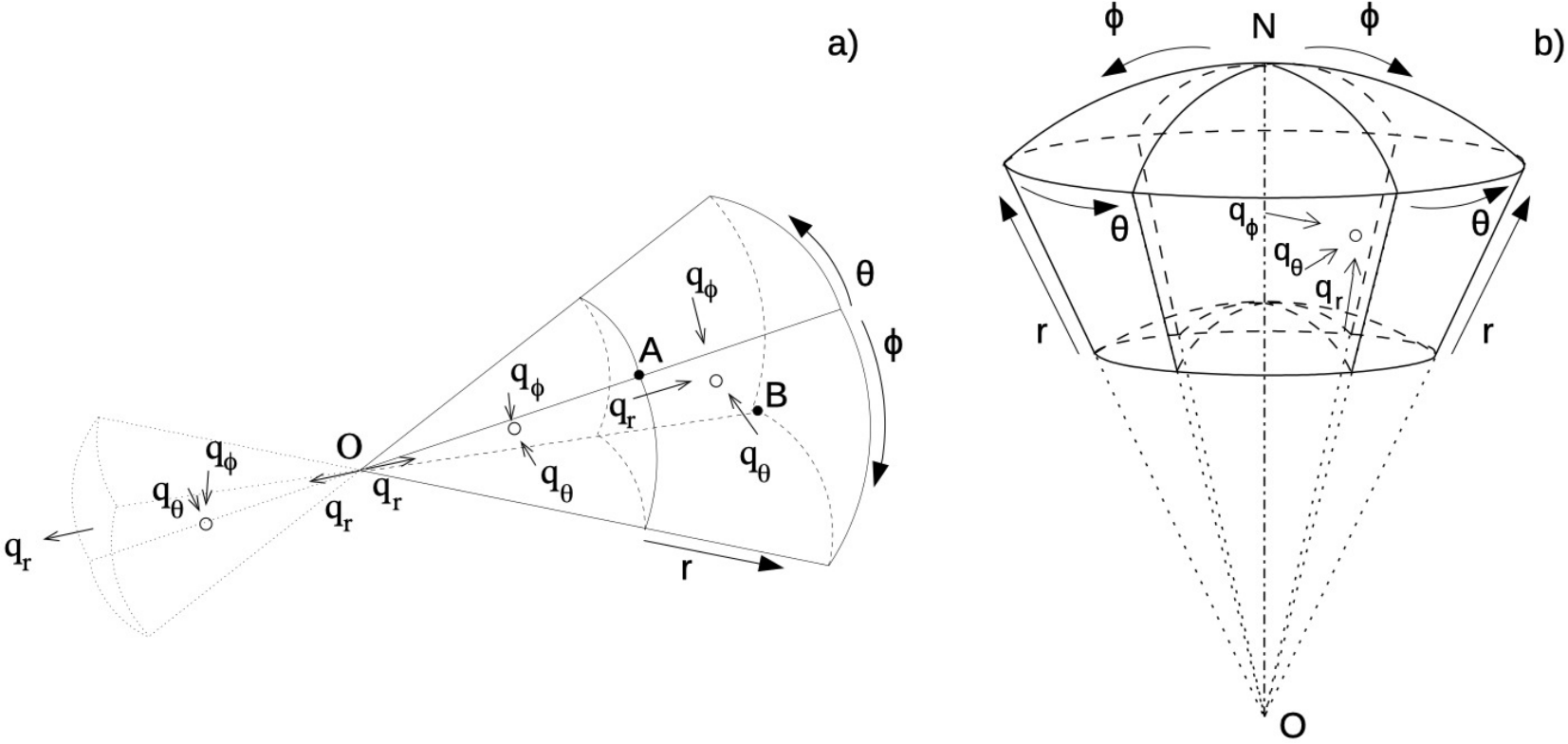}
\end{center}
\caption{a) Staggered arrangement of the discrete variables and cells next to the sphere
centre. b) Cells near the North pole of the sphere.}
\label{fig:2}
\end{figure}

We refer to the sketch of
Fig. \ref{fig:2} where the staggered arrangement of \cite{Harlo65} is adopted and
the node indices $1 \leq i\leq N_i$, $1 \leq j\leq N_j$, $1 \leq k\leq N_k$ span the 
$0 \leq \theta \leq 2 \pi$, $0 \leq r \leq R$ and $0 \leq \phi \leq \pi$ coordinates, 
respectively.

Let the nodes {\bf A} and {\bf B}
have, respectively, $i,j,k$ and
$i+1,j+1,k+1$ indices then $q_r(i,j,k)$ is located at the position
$(\theta_{i+1/2},r_j,\phi_{k+1/2})$ which is the centre of the $r$--normal
face of the cell. Similarly  $q_\theta(i,j,k)$ is at 
$(\theta_i,r_{j+1/2},\phi_{k+1/2})$, $q_\phi(i,j,k)$
at $(\theta_{i+1/2},r_{j+1/2},\phi_k)$ and the pressure at the cell centre
$(\theta_{i+1/2},r_{j+1/2},\phi_{k+1/2})$.
This implies that only the variable $q_r$ has $N_j$ values in the radial direction while
it has $N_i-1$ and $N_k-1$ values in the longitude and colatitude directions, respectively.
Similar considerations apply to the other velocity components.

Within the staggered discretization only $q_r$ is located at the sphere origin $r_1=0$
(Fig. \ref{fig:2}a) where, however, Equation (\ref{eq:qr}) does not 
need to be solved because $q_r(i,1,k) \equiv 0$ is a boundary condition.

The same argument applies to Equation (\ref{eq:qph}) in which, for  
$\phi = 0$ and 
$\phi = \pi$, it results $q_\phi \equiv 0$.

The variable change combined with the staggered discretization allows the straightforward
computation of {\it almost} all terms of Equations (\ref{eq:conq}, \ref{eq:qth}--\ref{eq:qph}) without 
coping with the singularities. 
As an example, we take the term $(1/r^2)\partial (q_r q_\phi)/\partial r$ of the 
$q_\phi$ equation for which, because of the staggering, the first radial node is at $j=3/2$. 
Evidencing by bold face the indices in the differentiated direction we obtain:

\setcounter{equation}{4}
\begin{equation}
\left .
\frac{1}{r^2}\frac{\partial q_r q_\phi }{\partial r} \right |_{i+\frac{1}{2}, {\bf \frac{3}{2}},k} \approx
\frac{(q_r q_\phi)_{i+\frac{1}{2},{\bf 2},k} -(q_rq_\phi)_{i+\frac{1}{2},{\bf 1},k}}{r^2_{\bf \frac{3}{2}}(r_{\bf 2}-r_{\bf 1})}.
\end{equation}
While we have omitted all the 
 averages needed to locate the unknowns at the appropriate positions for the differentiation,
(e.g. $(q_r q_\phi)_{i+1/2,2,k}= [q_r(i,2,k+1)+q_r(i,2,k)][q_\phi(i,3,k)+q_\phi(i,2,k)]/4$)
we note that all the quantities are perfectly defined
($r_{3/2} = \Delta r_1/2$ and $(q_r)_{i+1/2,1,k} =0$) and the derivative can be computed 
without any problem.

Along the same line, the term $1/(r \sin \phi) \partial (q_\theta q_\phi )/\partial \phi$ of the
$q_\theta$ equation computed at the North pole ($k=3/2$) node yields:
\begin{equation}
\left .
\frac{1}{r\sin \phi}\frac{\partial q_\phi q_\theta }{\partial \phi} \right |_{i,j+\frac{1}{2},{\bf \frac{3}{2}}} \approx
\frac{(q_\theta q_\phi)_{i,j+\frac{1}{2},{\bf 2}} -(q_\theta q_\phi)_{i,j+\frac{1}{2},{\bf 1}}}{r_{j+\frac{1}{2}}\sin \phi_{\bf \frac{3}{2}}(\phi_{\bf 2}-\phi_{\bf 1})},
\end{equation}
that, again, is not singular neither at the North pole ($\sin \phi_{3/2}\neq 0$) 
nor at the first radial node ($r_{3/2}\neq 0$).

The evaluation of the viscous terms at the origin ($r=0$) and at the poles
($\phi=0$ and $\phi=\pi$) benefits from the presence of the metrics that
avoid the explicit computation of derivatives at these locations. For example,
the discrete term
$(1/r^2)\partial /\partial r ( r^2\partial q_\theta / \partial r)$ in the $q_\theta$ 
equation at $j=3/2$ is:
\begin{equation}
\left .
\frac{1}{r^2}\frac{\partial }{\partial r} r^2\frac{\partial q_\theta}{\partial r} \right
|_{i,{\bf \frac{3}{2}}, k+\frac{1}{2}} \approx
\end{equation}
$$
\frac{1}{r_{\bf \frac{3}{2}}^2 }\left [ \left ( r^2\frac{\partial q_\theta}{\partial r} 
\right )_{i,{\bf 2},k+\frac{1}{2}} - \left ( r^2\frac{\partial q_\theta}{\partial r} 
\right )_{i,{\bf 1},k+\frac{1}{2}} \right ]/(r_{\bf 2}-r_{\bf 1}),
$$
that does not need the evaluation of $\partial q_\theta /\partial r$ at $j=1$ 
being multiplied by $r_1 \equiv 0$.

Similarly, the term $1/(r^2 \sin \phi) \partial /\partial \phi (\sin \phi \partial q_r/\partial \phi)
$ of the $q_r$ equation at the North pole reads:
\begin{equation}
\left .
\frac{1}{r^2 \sin \phi}\frac{\partial }{\partial \phi}\sin \phi \frac{\partial q_r}{\partial \phi}
\right |_{i+\frac{1}{2},j,{\bf \frac{3}{2}}} \approx
\end{equation}
$$
\frac{1}{r_j^2 \sin \phi_{\bf \frac{3}{2}}} \left [ \left ( \sin \phi \frac{\partial q_r}{\partial \phi } \right )_{i+\frac{1}{2},j,{\bf 2}}- \left (
\sin \phi \frac{\partial q_r}{\partial \phi } \right )_{i+\frac{1}{2},j,{\bf 1}} \right ]/(\phi_{\bf 2}-\phi_{\bf 1})
$$
and it does not need the evaluation of $\partial q_r/\partial \phi$ at $k=1$ since $\sin \phi_1 
\equiv 0$. Note that the term $1/r^2$ is not a problem because the $q_r$ equation is evaluated
only for $j \geq 2$.

Despite the change of variables and the staggered discretization still there are 
few terms that need a special treatment at the singular points; one of them 
is 
$\partial/\partial r [r^2 \partial (q_r/r^2)/\partial r]$ of the $q_r$ equation that
for $j=2$ becomes 
$$
\left .
\frac{\partial }{\partial r} r^2\frac{\partial q_r/r^2}{\partial r} 
 \right |_{i+\frac{1}{2},{\bf 2},k+\frac{1}{2}} \approx
\left \{ r^2_{\bf \frac{5}{2}} 
 \left [ \frac{(q_r)_{i+\frac{1}{2},{\bf 3},k+\frac{1}{2}}}{r_{\bf 3}^2}- \frac{(q_r)_{i+\frac{1}{2},{\bf 2},k+\frac{1}{2}}}{r_{\bf 2}^2} \right ]/(r_{\bf 3}-r_{\bf 2})
\right .
$$
\begin{equation}
\left .
- r^2_{\bf \frac{3}{2}}
 \left [ \frac{(q_r)_{i+\frac{1}{2},{\bf 2},k+\frac{1}{2}}}{r_{\bf 2}^2}- \frac{(q_r)_{i+\frac{1}{2},{\bf 1},k+\frac{1}{2}}}{r_{\bf 1}^2} \right ]/(r_{\bf 2}-r_{\bf 1})
\right \}/(r_{\bf \frac{5}{2}}-r_{\bf \frac{3}{2}}),
\label{eq:tedi}
\end{equation}
with the quantity $(q_r)_{i+1/2,1,k+1/2}/r^2_1$ that can not be evaluated directly. 

A possible strategy 
is to transform the derivative 
\begin{equation}
\frac{\partial }{\partial r} r^2\frac{\partial q_r/r^2}{\partial r} \equiv
\frac{\partial^2 q_r}{\partial r^2} - \frac{\partial}{\partial r} \left ( \frac{2 q_r}{r}\right ),
\end{equation}
whose right hand side that can be computed in a straightforward way. 
It is worth mentioning, however, that the above equivalence 
is valid only in the continuum limit while differences arise when both 
sides are discretized.

A different approach is to maintain the formulation (\ref{eq:tedi}) and replace the 
singular quantity by a surrogate obtained by an average with the two
counterparts `opposite' to the 
singular point. This approach had already been used successfully 
for the axis of polar cylindrical 
coordinates by \cite{Eggel94} and \cite{Verzi96} although it has  to be modified in 
the spherical case since the singularities at the centre and at the polar axis need
a different treatment.

In the case of Equation (\ref{eq:tedi}) we recall that $q_r/r^2 = u_r$ and use the
second--order midpoint interpolation $u_r(\theta, 0,\phi) = (u_r(\theta, \Delta r, \phi)-
u_r(\theta+\pi, \Delta r, \pi-\phi))/2 + {\cal O}(\Delta r^2)$ in which the minus comes
from the opposite orientation of the radial velocity in the mirror plane 
(Fig. \ref{fig:2}a).
With the discrete variables we have
\begin{equation}
\frac{q_r(i,1,k)}{r_1^2} \approx \frac{1}{2} 
\left [ \frac{q_r(i,2,k)}{r_2^2} - \frac{q_r(i+N_{ir}/2,2,N_{kr}-k)}{r_2^2} \right ],
\end{equation}
with $N_{ir}=N_i-1$ and $N_{kr}=N_k-1$ the number of $q_r$ points in the azimuthal and
colatitude directions,
that make possible the evaluation of the viscous term through (\ref{eq:tedi}).

All the other terms needing the evaluation at $j=1$ of $q_r/r^2$ can be treated in
the same way.

Similarly, for the $q_\phi$ equation, the term 
$(1/r)\partial (q_\phi q_\phi/\sin \phi)$ 
requires the evaluation of $q_\phi/\sin \phi$ at $k=1$ and $k=N_k$ which can not be done 
directly. Again we use the relation $q_\phi/\sin \phi = u_\phi$ and write
$u_\phi(\theta, r,0) = (u_\phi(\theta, r, \Delta \phi)-
u_\phi(\theta+\pi, r, \Delta \phi))/2 + {\cal O}(\Delta \phi^2)$
that with the discrete variables becomes:
\begin{equation}
\frac{q_\phi(i,j,1)}{\sin \phi_1} \approx \frac{1}{2} 
\left [ \frac{q_\phi(i,j,2)}{\sin \phi_2} - \frac{q_\phi(i+N_{i\phi}/2,j,2)}{\sin \phi_2} \right ],
\end{equation}
with $N_{i\phi}=N_1-1$ and a similar expression for $\phi=\pi$ ($k=N_k$). The same approach can be adopted
for the viscous term needing the evaluation of $q_\phi/\sin \phi$ at the polar axis.

Note that, differently form the sphere centre, at the polar axis there is no colatitude 
inversion to select the mirror point.

\textcolor{black}{
A possible cause of concern is that the special discretizations of some 
singular terms couple the meridional plane at $\theta_i$ with that 
at $\theta_i + \pi$ thus creating communication issues which would impede 
performance on highly parallel computers. This is however not the case
since the above procedures require only one extra halo cell at the singular
point and the involved data communication is irrelevant. In fact the same
strategy is used in \cite{Steve10} for the equations in cylindrical coordinates
that are solved on massive parallel computers using up to $3.2\times 10^4$
processors. 
}

\section{Numerical Method}
\label{sec:nm}

Equations (\ref{eq:conq}, \ref{eq:qth}--\ref{eq:qph}) are integrated using the fractional--step
method detailed in \cite{Verzi96} and briefly summarized below.
The momentum equation for ${\bf q}=(q_\theta, q_r,q_\phi)$ is provisionally advanced
in time using the old pressure field:
\begin{equation}
\frac{\widehat{{\bf q}}-{\bf q}^l}{\Delta t}= \left [ \gamma^l {\bf N}^l
+\rho^l {\bf N}^{l-1} -\alpha^l \nabla p^l + \alpha^l \frac{\widehat{{\bf V}}-{\bf V}^l}{2}\right ].
\label{eq:pmo}
\end{equation}
 Here the superscript $l$ indicates the time--step level, $\alpha^l$, $\gamma^l$ and
$\rho^l$ are the coefficients of the time integration scheme (second--order Adams--Bashfort or
third--order Runge--Kutta). ${\bf N}$ contains the 
explicit nonlinear terms, body forces and the off--diagonal viscous terms, while
${\bf V}$ the implicit diagonal viscous terms.

Since the pressure is not updated, the resulting velocity field is not 
locally free--divergent
and is denoted by $\widehat{{\bf q}}$. The correct velocity, however, must differ
from the provisional one only by a gradient term, thus we can write:
\begin{equation}
{\bf q}^{l+1} = \widehat{{\bf q}}-\alpha^l \Delta t \nabla \Phi,
\label{eq:vco}
\end{equation}
whose divergence yields the elliptic equation for the unknown correction:
\begin{equation}
\nabla^2 \Phi = \frac{\nabla \cdot \widehat{{\bf q}}}{\alpha^l \Delta t}.
\label{eq:ell}
\end{equation}
Once the scalar field $\Phi$ is determined, the solenoidal velocity ${{\bf q}}^{l+1}$
is computed by Equation (\ref{eq:vco}) and the new pressure through:
\begin{equation}
p^{l+1} = p^l + \Phi -\frac{\alpha^l \Delta t}{2Re} \nabla^2 \Phi.
\label{eq:npre}
\end{equation}

The implicit treatment of the diagonal viscous terms of (\ref{eq:pmo}) would require the inversion of 
a large sparse matrix that is very time consuming; this is avoided by using the approximate
factorization technique of \cite{Beam76} that requires only the inversion of three tridiagonal 
matrices with an error ${\cal O}(\Delta t^3)$. 

Being the nonlinear convective terms computed explicitly, the equations should
satisfy only the $CFL={\rm max}[ \Delta t (|r \sin \phi \Delta \theta/u_\theta|+|\Delta r/u_r| + 
|r\Delta \phi/u_\phi| )]$ stability conditions that is $CFL \leq 1$ for the Adams--Bashfort and
$CFL \leq \sqrt{3}$ for the third--order Runge--Kutta scheme. The off--diagonal viscous terms, however,
are also computed explicitly to avoid the implicit coupling of the three momentum equations
and this deteriorates the stability properties of the scheme.
 The actual $CFL$ value used for the simulations therefore
must be reduced with respect to the theoretical value and the amount of reduction depends on the
Reynolds number and on the specific flow. In our applications we have found that halving the convective
$CFL$ limit yields a safe enough condition that allows the stable integration of the equations; 
it is worth mentioning, however, that 
this criterion must be taken as a rule of thumb and not as a strict limitation.

The current implementation of the method allows the use of generic non--uniform mesh distributions
in the radial and colatitude directions. The reason for maintaining the uniform discretization in 
the remaining  longitudinal 
direction is that 
we use trigonometric
expansions and fast--Fourier--transforms to reduce the elliptic Equation (\ref{eq:ell}) 
to a series
of two--dimensional Helmholtz equations
in the other two coordinates that are solved using the direct method of \cite{Schwa74} or \cite{Lynch64}.

\section{Results}
\label{sec:re}

In this Section we assess the qualities of the proposed numerical method by
showing several numerical examples and benchmarking them with the results
from other codes or those available from the literature.

\subsection{Hill vortex}
\label{sec:HV}

As a first example we consider a spherical Hill vortex \cite{Hill94} which
is an exact solution of the Euler equations and, in the inviscid limit, preserves
its shape and propagates with a constant velocity along a rectilinear trajectory.

\begin{figure}[htb!]
\begin{center}
\includegraphics[width= 0.95 \textwidth]{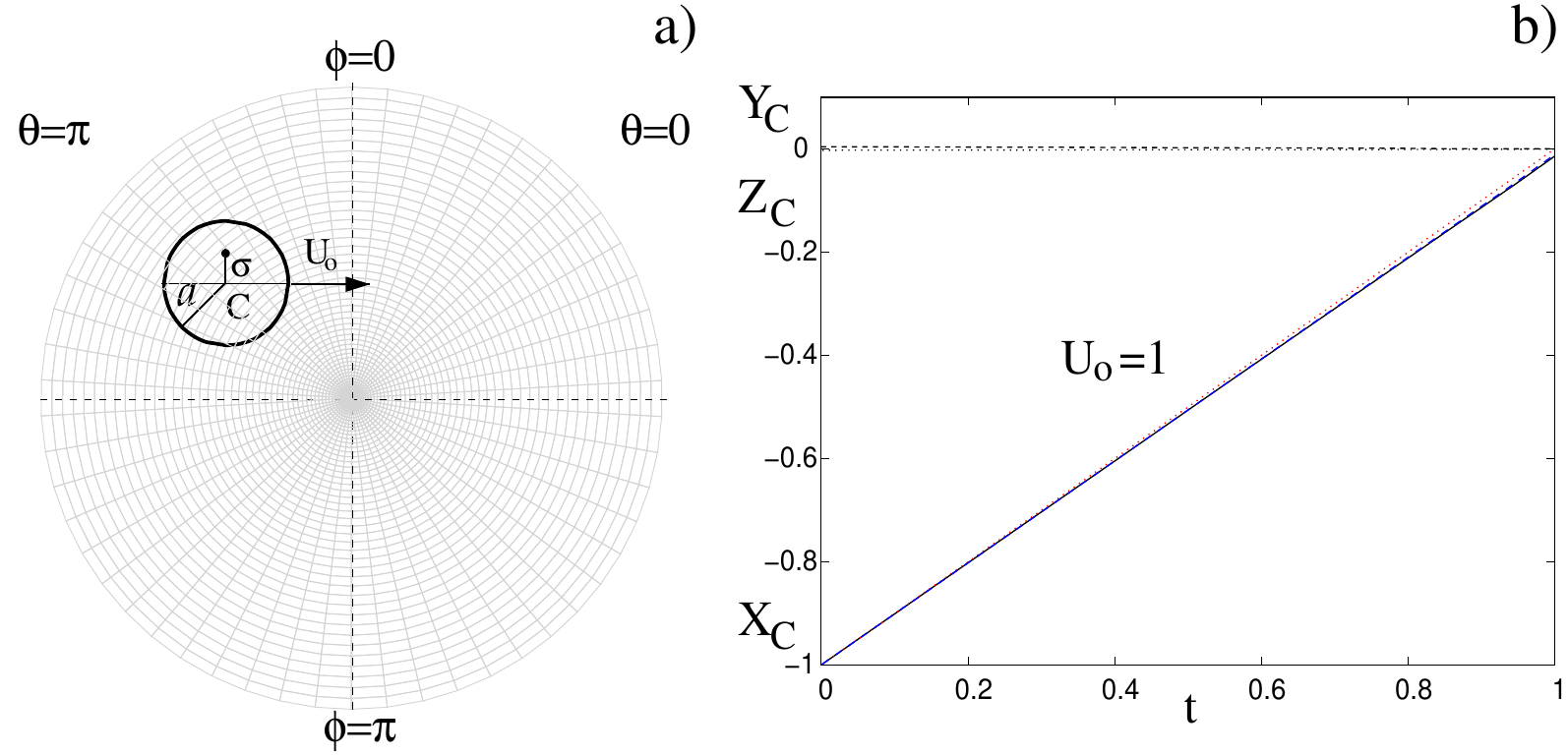}
\end{center}
\caption{a) Sketch of the Hill vortex setup in the longitudinal sections
          $\theta=0,\pi$ of the spherical domain.
         b) Time evolution of the vortex centre coordinates 
for the case at $Re=2500$ with
$\theta_C = \pi$, $r_C=1$, $\phi_C=\pi/2$.
\solid $X_C = r_C \cos \theta_C \sin \phi_C$, {\color{black} \dashed} 
$X_C$ from the code in Cartesian coordinates \cite{Afid}, thin {\color{red}\dotted } theoretical 
$X_C$; \dashed $Y_C = r_C \sin \theta_C \sin \phi_C$; \chndot
$Z_C = r_C \cos \phi_C$.
}
\label{fig:H0}
\end{figure}

The vortex is defined by assigning its toroidal vorticity
$\omega = A\sigma$ within the sphere of radius $a$ and centre
$C$ with an irrotational flow outside; the translation velocity
of the ring is then $U_0 = 2Aa^2/15$ (Fig. \ref{fig:H0}a) and the Reynolds number
is defined as $Re=2aU_0/\nu$. 
Different initial positions $C$ and velocity orientations $U_0$ have been
simulated in order to stress the stability, the accuracy  and the reliability of the method.

If not specified otherwise, the simulations have been performed in a domain of radius
$R=7a$ discretized by a mesh of $129^3$ nodes using $Re=2500$.
 
\begin{figure}[htb!]
\begin{center}
\includegraphics[width= 0.98 \textwidth]{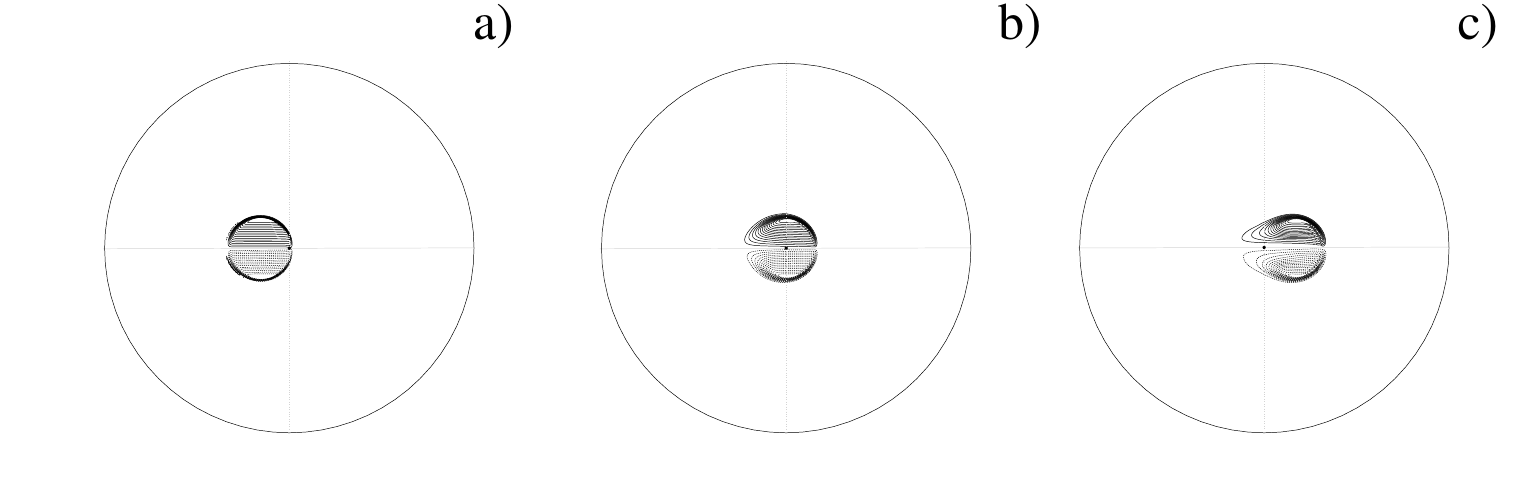}
\end{center}
\caption{Time evolution of the Hill vortex at $Re=2500$
in the longitudinal sections 
          $\theta=0,\pi$; longitudinal vorticity
($\Delta \omega_\theta= \pm 0.4$) 
\dotted for negative values;
a) ) $t=0.5$, b) $t=1.$, c) $t=1.5$. 
Initial centre of the vortex: $\theta_C = \pi$, $r_C=1$, $\phi_C=\pi/2$.
Hereinafter, the black bullet indicates the origin of the spherical system.}
\label{fig:H11}
\end{figure}

\begin{figure}[htb!]
\begin{center}
\includegraphics[width= 0.98 \textwidth]{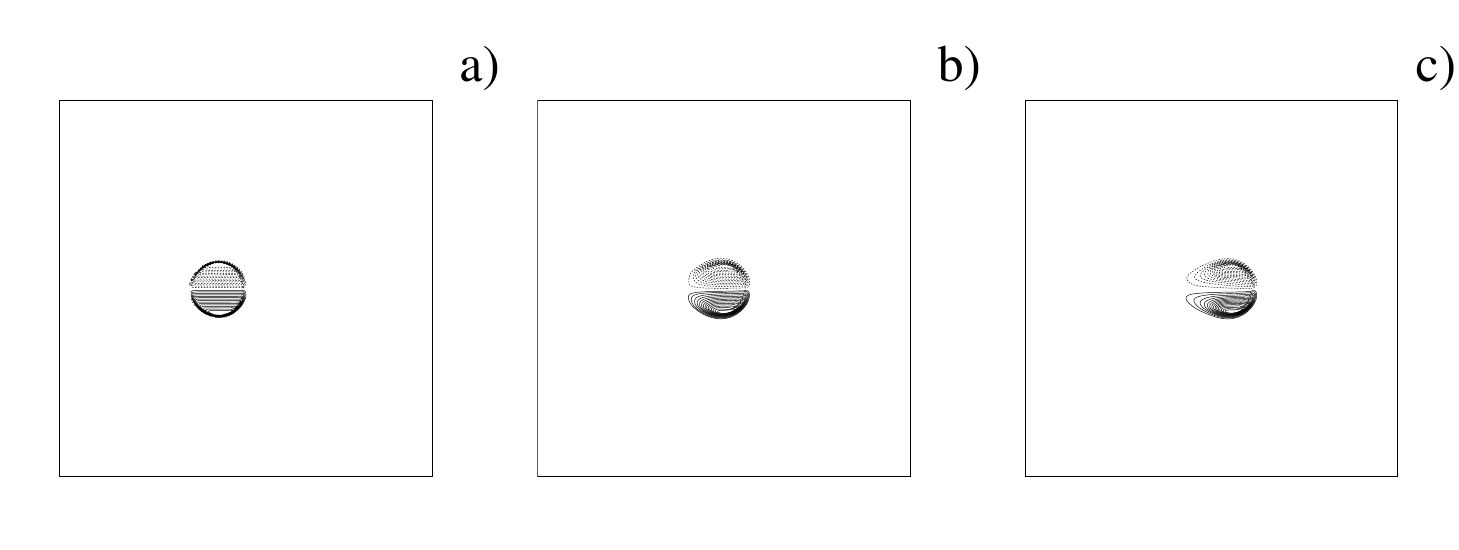}
\end{center}
\caption{Time evolution of the Hill vortex at $Re=2500$ evolved on
a Cartesian uniform mesh; 
          out--of--the--page vorticity
($\Delta \omega = \pm 0.4$) 
\dotted for negative values;
a) ) $t=0.5$, b) $t=1.$, c) $t=1.5$.}
\label{fig:HCa}
\end{figure}

In Fig. \ref{fig:H0}b we report the trajectory of the vortex, through the
Cartesian coordinates of the velocity peak, that moves horizontally with a constant velocity and 
crosses the sphere centre ($r=0$) (Fig. \ref{fig:H11}). 
It can be noted that while $Y_C$ and $Z_C$ remain negligible in time, $X_C$ increases
linearly thus confirming the constant translation velocity. In fact, a close inspection
of $X_C$ reveals a small deviation with respect to the theoretical straight line; this
is due to the finite viscosity of the flow ($Re=2500$) that perturbs the exact inviscid 
solution and deforms the initial vortex shape. That the vortex deformation is indeed
due to viscosity and not to the discretization on spherical coordinates is confirmed by the results of Fig. \ref{fig:HCa} in which the same Hill vortex has been evolved
using the code in Cartesian coordinates AFiD \cite{Afid}. From the same calculation we have
extracted also the trajectory of the vortex that perfectly overlaps with its counterpart
computed in spherical coordinates.

\begin{figure}[htb!]
\begin{center} 
\includegraphics[width= 0.95 \textwidth]{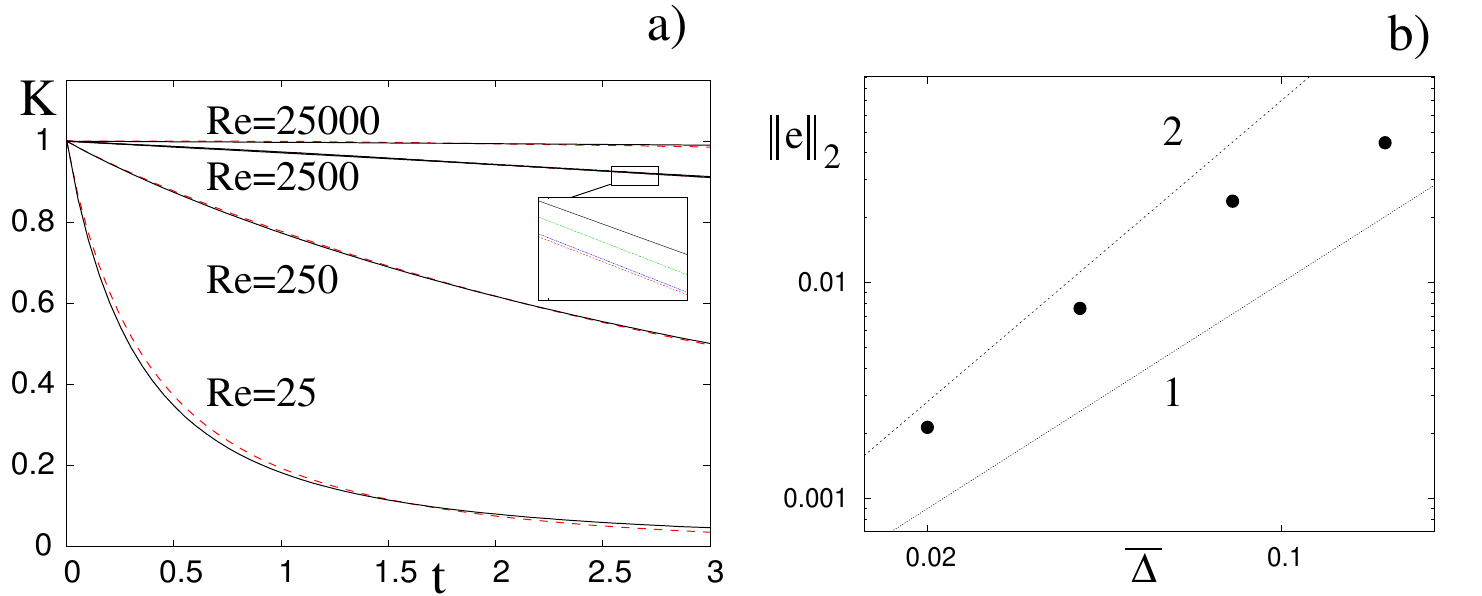}
\end{center}
\caption{a) Decay of the total kinetic energy in time for the Hill
vortex: \solid results from the code in spherical coordinates
{\color{red} \dashed } results from the Cartesian code \cite{Afid}.
In the inset:
{\color{blue} \chndash } and $Re=2500$ for the Hill vortex with initial offset.
{\color{green} \chndashdash } and $Re=2500$ for the Hill vortex resolved on a mesh refined around the singular points.
         b) $L_2$--norm of the error as function of the mesh size
for the case at $Re=2500$. \dashed $2$--slope, \dotted $1$--slope.
\textcolor{black}{$\overline{\Delta}$ is the mean grid spacing defined 
as $\overline{\Delta} \equiv [V/(N_iN_jN_k)]^{1/3}$ with $V$ the
volume of the computational domain.}
$\theta_C = \pi$, $r_C=1$, $\phi_C=\pi/2$.
}
\label{fig:H3}
\end{figure}

The same comparison between spherical and Cartesian codes has been repeated for a wide
range of Reynolds numbers and in Fig. \ref{fig:H3}a we report the time evolution of 
the flow kinetic energy ($K=0.5\int_V {\bf u}^2{\rm d}V$), normalized by the initial 
value, for several Reynolds numbers
showing that, despite the very different meshes, the kinetic energy decays always at
an identical rate that depends only on $Re$.

Even if the results are the same,
the simulations in spherical coordinates are more expensive than the
Cartesian counterparts owing to the time step limitations introduced by the
mesh around $r=0$. For example, using the third--order Runge--Kutta as time integration
scheme and a working $CFL$ of $1.2$, the simulation at $Re=2500$ on a $129^3$ mesh in a 
$4^3$ domain run with a time step $\Delta t\simeq 2\times 10^{-2}$ throughout the whole 
computation. The same case on the spherical mesh had $CFL=0.6$
with a dynamically adjusted time step of $\Delta t \simeq 10^{-2}$, when the
vortex was far from the sphere centre, and $\Delta t \simeq 10^{-4}$ during the
crossing phase. As a result, the CPU time for the latter simulation was about
$14$ times bigger than that of the former.

We wish to point out that the reason for this large computational overhead is the Hill vortex flow that,
with its strongest velocity components perpendicular to the polar axis 
\textcolor{black}{near the sphere centre}, is 
particularly unsuitable for the spherical discretization. 
Nevertheless it has been chosen on purpose, in order to show that the numerical method can be
used even in the most unfavourable conditions without loosing
stability or precision. 
\textcolor{black}{It is important to note that the time step 
limitation is mainly given by the discretization at the sphere centre and
not by the polar axis. In fact, later in this section we show that 
when the Hill vortex has an initial offset of $\phi_C=2\pi/3$,
with respect to the symmetry plane $\phi = \pi/2$, the time step reduction
during the crossing of the polar axis is only a factor $\approx 3$--$4$ even
if the largest velocity is still perpendicular to the coordinate line
$\phi=0$.
Finally, in the next sections we will consider numerical examples in which 
the flow evolves in between two spherical shells and there the time step
is limited only by the radial refinement of the mesh at the solid 
boundaries and not by the singularity at the polar axis.}

Since the most critical phase of the simulation is the vortex centre crossing the
point $r=0$, here we have evaluated the code accuracy by assuming as reference
solution that performed on the finest mesh $1459\times 1297\times 730$ 
(in $\theta$, $r$ and $\phi$) and 
comparing it with grids successively coarsened by a factor $3$ in each direction. 
This coarsening factor is such that, on a staggered mesh, it allows to compare the
velocity components of different grids without interpolation  and therefore
to compute the raw accuracy of the numerical method. 
This set of simulations has been run with the same constant time step 
($\Delta t = 6\times10^{-6}$) that was imposed by the stability of the
simulation run on the finest mesh. 

In Fig. \ref{fig:H3}b we
report the $L_2$--norm of the error computed for four meshes and, apart
for the coarsest ($19\times 17\times 10$), the error decreases quadratically with 
the mesh size thus confirming the second--order accuracy.

\begin{figure}[htb!]
\begin{center}
\includegraphics[width= 0.98 \textwidth]{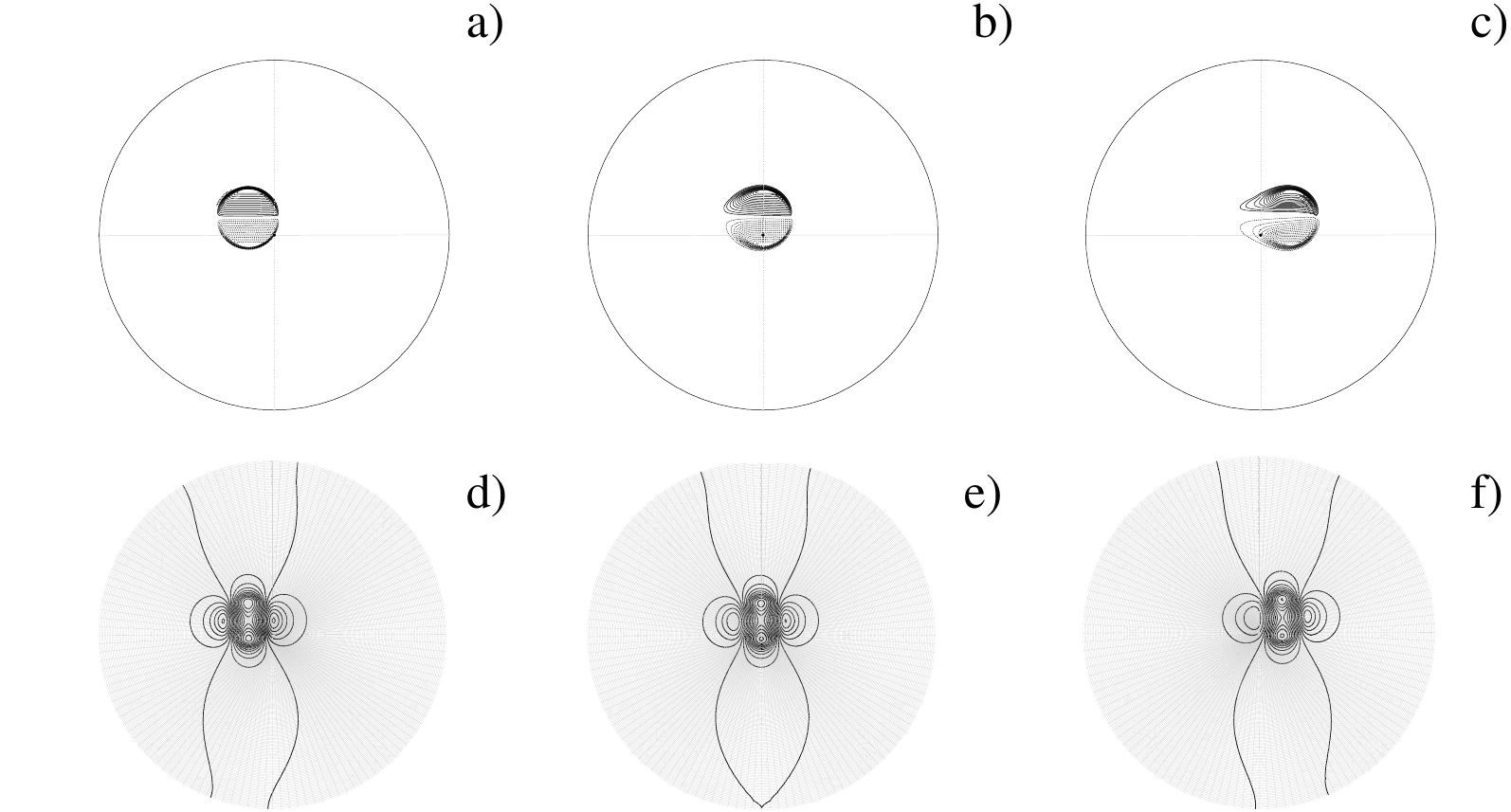}
\end{center}
\caption{Time evolution of the Hill vortex at $Re=2500$ in the longitudinal sections
          $\theta=0,\pi$; top panels show longitudinal vorticity
($\Delta \omega_\theta= \pm 0.4$) 
\dotted for negative values;
($\Delta u=0.1$) bottom panels for pressure ($\Delta p = 0.02$); the background
mesh is shown with light gray lines.
a) and d) $t=0.5$, b) and e) $t=1.$, c) and f) $t=1.5$. 
Initial centre of the vortex: $\theta_C = \pi$, $r_C=1$, $\phi_C=2\pi/3$.}
\label{fig:Hoff}
\end{figure}

We have further stressed the numerical method by giving the Hill vortex an initial
offset with respect to the symmetry plane $\phi=\pi/2$.
It can be observed from Fig. \ref{fig:Hoff} that in this 
case no coordinate lines are aligned with the vortex axis and nevertheless the vortex
translates along a horizontal rectilinear trajectory showing the same dynamics as
in Fig. \ref{fig:H11}. Also the total kinetic energy of the flow decays in time in the same
way as the other $Re=2500$ cases (Fig. \ref{fig:H3}a) and this confirms that 
the vortex evolution does not depend on the mesh orientation.
\textcolor{black}{
Similarly to the case of Fig. \ref{fig:H11}, also for this simulation
the time step has been dynamically computed in time to keep the $CFL$ 
constant at the value of $0.6$. In this case, however, it resulted
$\Delta t \simeq 10^{-2}$ at the beginning of the simulation and it decreased
to $\Delta t \simeq 2.7\times 10^{-3}$ during the crossing of the polar axis.
Considering that for a Hill vortex the largest velocity occurs at the centre,
comparing Figs. \ref{fig:H11}b and \ref{fig:Hoff}b, it is clear that the most 
important limitation to the time step comes from the singularity around the 
sphere centre and not by that at the polar axis.
}

\begin{figure}[htb!]
\begin{center}
\includegraphics[width= 0.98 \textwidth]{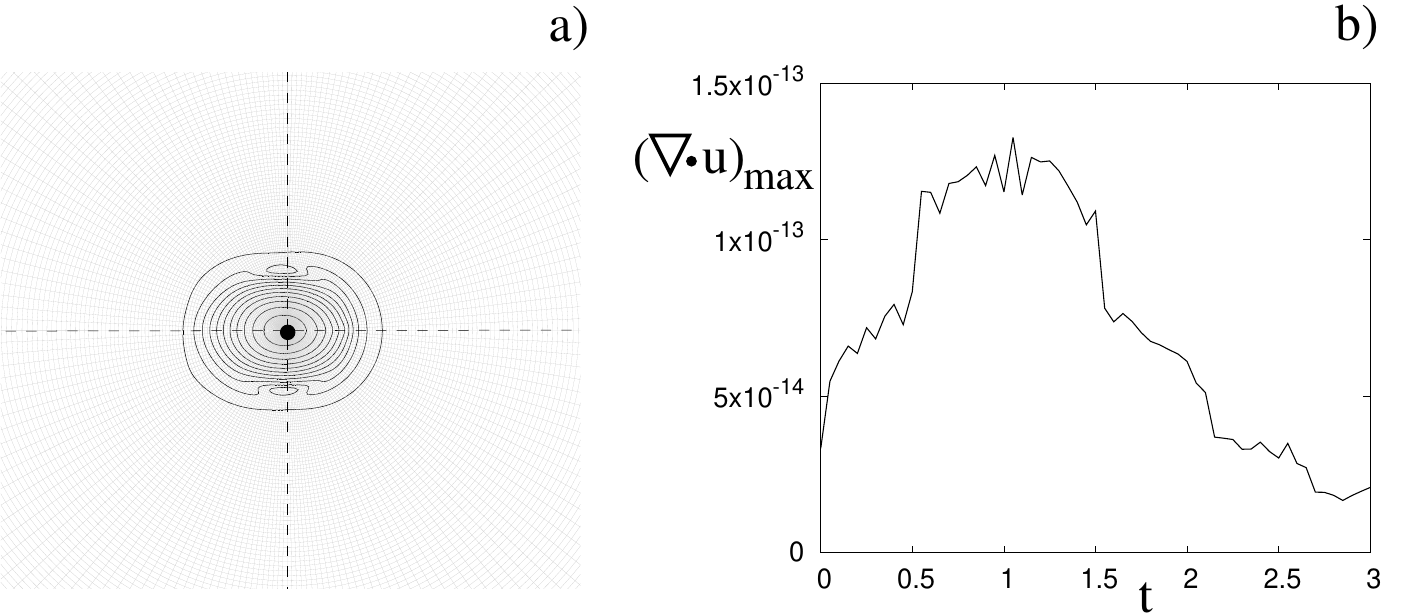}
\end{center}
\caption{a) Zoom of velocity magnitude ($\Delta u=0.1$) of a Hill vortex at $Re=2500$ 
and $t=1$ in the longitudinal sections
          $\theta=0,\pi$; the computational mesh is refined on purpose
around the singular points.
b) Time evolution of the maximum divergence of the velocity field.
Initial centre of the vortex: $\theta_C = \pi$, $r_C=1$, $\phi_C=\pi/2$.}
\label{fig:HL}
\end{figure}

In another test we have positioned the vortex as in the case of Fig. \ref{fig:H11} but
the mesh has been refined at the sphere centre and at the polar axis (in both cases 
using a hyperbolic tangent distribution with stretching parameter $1.5$) in order to 
exacerbate the stability problems of the integration in spherical coordinates.
The results of Fig. \ref{fig:HL}a show that also in this case the vortex crosses the singular region 
without being distorted and the maximum divergence of the flowfield remains at machine 
precision throughout all the evolution (Fig. \ref{fig:HL}b). 
Also the total kinetic energy decay is identical
to those of the other $Re=2500$ cases (Fig. \ref{fig:H3}a) again showing a vortex
dynamics independent of the mesh distribution.

\textcolor{black}{
\subsection{Flow in a precessing and spinning sphere}
\label{sec:PSS}
\cite{Auti09} validated their pseudo--spectral method by replicating the
same problem as in \cite{Kida08} who studied the flow in a 
precessing and spinning sphere. We benchmark our code using the same test case
and,
referring to the sketch of Fig. \ref{fig1n}, we consider flow inside a sphere of
radius $R$ spinning about the $x$--axis at constant angular velocity
$\Omega_s$. The system has an additional precession angular velocity
$\Omega_p$ about the $z$--axis and the sphere surface ($r=R$) is no--slip.
Assuming $R$ and $\Omega_sR$, respectively, as scaling length and velocity,
the flow depends on two nondimensional parameters
$Re = \Omega_s R^2/\nu$ and $\Gamma = \Omega_p/\Omega_s$.
}

\textcolor{black}{
Following \cite{Kida08}, \cite{Auti09} we solve the equations in the 
precessing reference frame 
\begin{equation}
\frac{\partial {\bf u}}{\partial t} + \nabla \cdot ({\bf u}{\bf u}) =
-\nabla P -2\Gamma \widehat{k}\wedge {\bf u} + \frac{1}{Re}\nabla^2 {\bf u},
\qquad \qquad \nabla \cdot {\bf u}=0,
\end{equation}
being $\widehat{k}$ the unit vector of the $z$--axis and 
$P = p -(\Gamma^2/2) (\widehat{k}\wedge {\bf r})^2$ the reduced pressure.
The boundary condition for the nondimensional velocity at the sphere 
surface is ${\bf u}|_{r=R} = \widehat{x}\wedge \widehat{r}$ with $\widehat{x}$
and $\widehat{r}$
the unit vectors of the $x$--axis and of the radius ${\bf r}$.
}

\textcolor{black}{
\cite{Kida08} analysed the flow for $\Gamma= 0.1$ and $Re\leq 500$ 
finding that a steady state is eventually achieved with  increasingly 
entangled toroidal structures developed within the sphere. 
On the other hand, \cite{Auti09}
observed that the flow structure could be better understood by plotting
the velocity field in a reference frame rotating with the spinning
sphere ${\bf v} = {\bf u} - \widehat{x}\wedge {\bf r}$.
}

\begin{figure}[htb!]
\begin{center}
\includegraphics[width= 0.78 \textwidth]{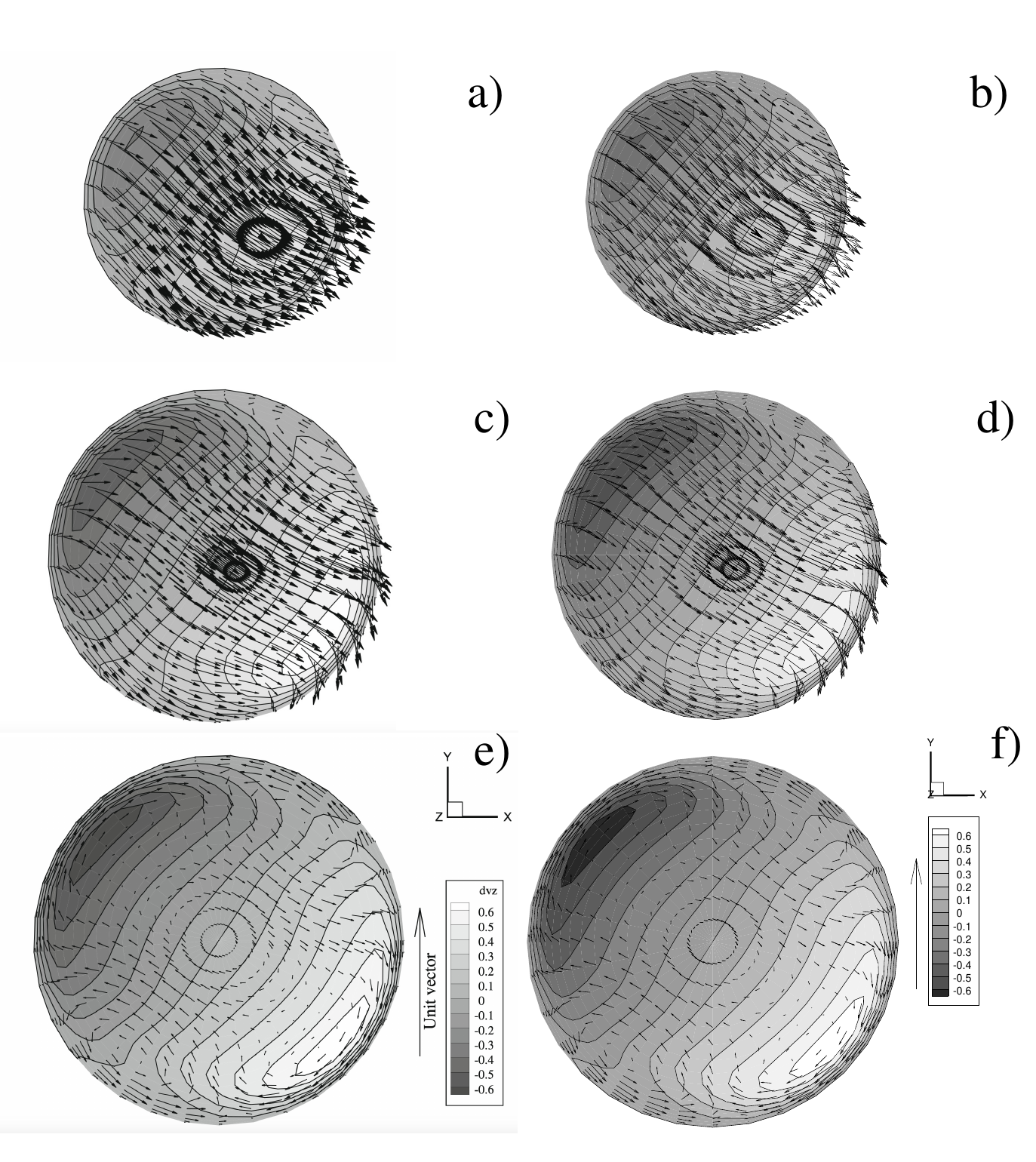}
\end{center}
\caption{
\textcolor{black}{
Flow in a precessing, spinning sphere at $\Gamma=0.1$ and 
$Re=500$. The velocity vectors of the field rotating with the spinning
sphere ${\bf v}$ are reported in the sections $z=-0.66$, a) and d);
$z=-0.33$, b) and e); $z=0$, c) and f).
The countours are for the velocity component $v_z$.
a), c) and e) are the results adapted from Fig. 7 of \cite{Auti09},
b) ,d) and f), present results.}
}
\label{fig:prec}
\end{figure}

\textcolor{black}{
Here we consider the case at the highest Reynolds number $Re=500$ and compare
our results for the steady state solution of ${\bf v}$ at the same three
$z=const$ planes as reported by \cite{Auti09}. The simulations have been
performed at the same resolution and the comparison,
 given in Fig. \ref{fig:prec}, shows excellent agreement.
}

\textcolor{black}{
Once again we note that in this numerical example, being the flow velocities
very small at the sphere centre (see Figs. \ref{fig:prec}cf) the integration could
be carried out at constant $CFL=0.6$ with a nondimensional time step of
$\approx 4\times 10^{-2}$ and the limitation was generated 
by the fine mesh stretched at the sphere surface rather than from the singular
points of the spherical coordinates.
}

\subsection{Rayleigh--B\'enard convection with vertical gravity}
\label{sec:RB}

\begin{figure}[htb!]
\begin{center} 
\includegraphics[width= 0.98 \textwidth]{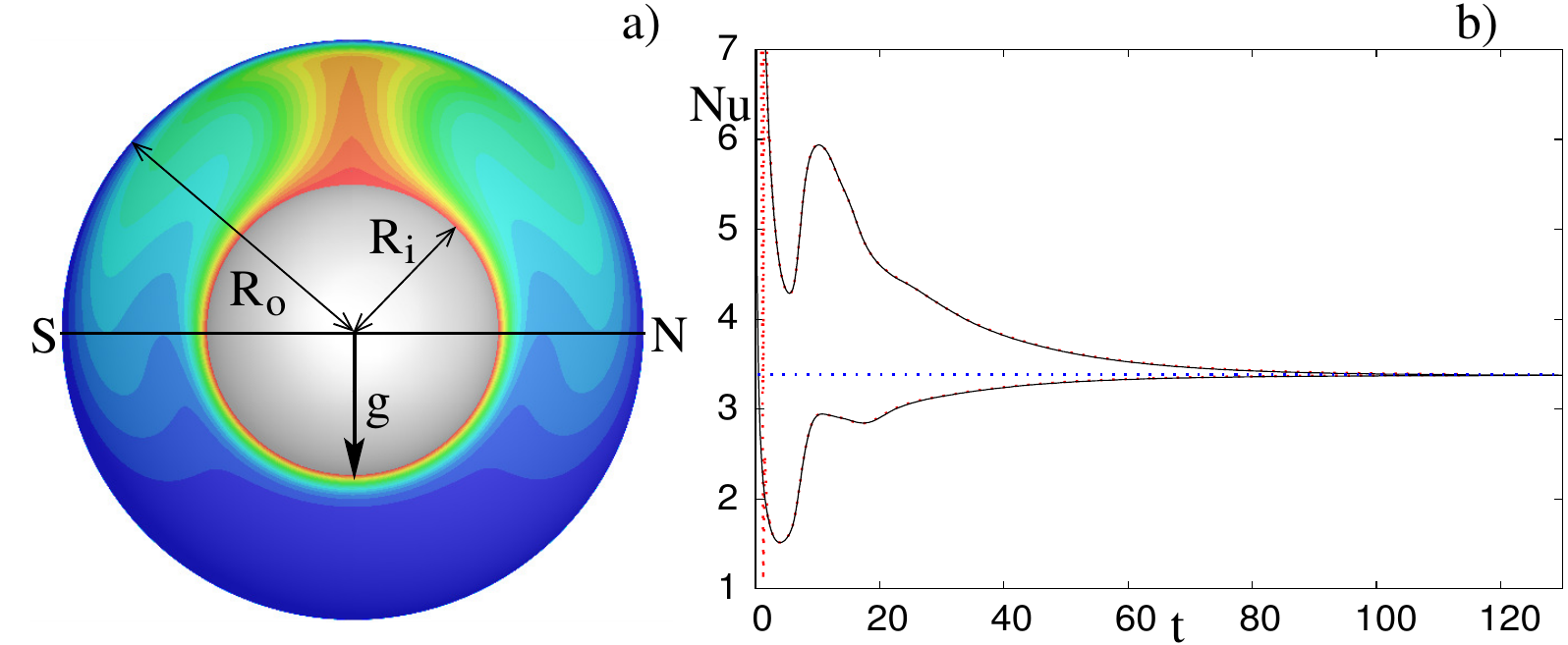}
\end{center}
\caption{a) Section through the meridional planes $\theta=0,\pi$ of temperature 
         contours ($\Delta T = 0.1$) for the flow at $\eta=R_i/R_o=0.5$,
$Ra=10^5$ and $Pr=0.7$. Grid $65\times 49\times 49$.
         b) Time evolution of the Nusselt numbers \solid simulation with the gravity
aligned with the polar axis, {\color{red}\dashed } gravity perpendicular to the polar axis,
{\color{blue} \dotted } reference value of $Nu=3.4012$ from \cite{Feldm13}.
}
\label{fig:RB1}
\end{figure}

In this example
 we consider the thermally driven flow developing between two concentric spheres
of radii $R_i$ and $R_o$  whose surfaces are maintained at a temperature difference
$\Delta T$, the inner being hotter than the outer. Following the arrangement of
\cite{Feldm13} the gravity has a constant orientation and points vertically downward
 as in Fig. \ref{fig:RB1}a. 

The flow is solved using the Boussinesq approximation in which all the fluid properties
are independent of the temperature except for the density in the buoyancy term of the
momentum equation. In addition to the conservation of mass and balance of momentum 
(Equations \ref{eq:ns})  here we need also the energy conservation that reduces to the convection--diffusion
equation for the temperature field. The complete set of governing relations in
non-dimensional vector form then reads:

$$
\nabla \cdot {\bf u} = 0,
$$
$$
\frac{\partial {\bf u}}{\partial t} + {\bf u}\cdot \nabla {\bf u} = -\nabla p
-\widehat{g}T +\sqrt{\frac{Pr}{Ra}} \nabla^2 {\bf u},
$$
\begin{equation}
\frac{\partial T}{\partial t} + {\bf u}\cdot \nabla T = \sqrt{\frac{1}{RaPr}}\nabla^2 T.
\label{eq:Bou}
\end{equation}

$Ra = g \beta \Delta t (R_o-R_i)^3/(\nu \kappa)$ is the Rayleigh number with $\beta$ the
isobaric thermal expansion coefficient, $\kappa$ the thermal diffusivity of the fluid,
$g$ the magnitude of the gravity and $\widehat{g}$ its unit vector;
$Pr=\nu/\kappa$ is the Prandtl number. 
Referring to Equations (\ref{eq:ns}) we can also write $\widehat{g}=
(f_\theta, f_r, f_\phi) = (\sin \theta, \sin \phi \cos \theta, \cos \phi \cos \theta$).

Being the temperature a scalar quantity it is located at the cell centre 
(Fig. \ref{fig:2}) and the solution of the last of Equations (\ref{eq:Bou})
in spherical coordinates does not present particular challenges at the singular points.

In the present flow, a buoyant plume is produced that ascends vertically thus the `natural' 
arrangement is to have the gravity vector aligned with the polar axis so that there are no
velocity vectors crossing it. 
\textcolor{black}{
In order to show that the proposed numerical 
method, performs well also in supposedly unfavourable conditions,
we have repeated the simulation 
also with the gravity perpendicular to the polar axis.
}

In Fig. \ref{fig:RB1} we report the results for a case at $\eta=R_i/R_o=0.5$,
$Ra=10^5$ and $Pr=0.7$ that, after an initial transient, attains a steady state.

A usual way to express the heat transfer in thermally driven flows is by the Nusselt 
number defined as the ratio between the heat flux through a surface and its counterpart
in absence of flow motion. For this problem it can be computed for the inner and
the outer spheres to obtain:
\begin{equation}
\left .
Nu_o = -\frac{1}{\eta} \overline{\frac{\partial T}{\partial r}} \right |_{r=R_i},
\qquad
\left .
Nu_i = -\eta \overline{\frac{\partial T}{\partial r}} \right |_{r=R_o},
\label{eq:Nu}
\end{equation}
where the $\overline{\ \cdot\ }$ indicates surface and time averages:
if the flow is steady or it attains a statistical steady state the two values have to
match.

Fig. \ref{fig:RB1}b shows the time evolution of the inner and outer Nusselt
numbers for the simulations with the gravity in two perpendicular orientations; it
can be observed that not only they converge exactly to the same value but also the
transient evolutions are indistinguishable. The asymptotic Nusselt number is 
$Nu=3.4105$ in excellent agreement with the values $Nu=3.4012$ of \cite{Feldm13}, $3.4890$ of \cite{Chu93}
$3.4648$ of \cite{Dehg10} and $3.3555$ of \cite{Chiu96}.

\textcolor{black}{
Another interesting result is that in both cases, with the gravity aligned or
perpendicular to the polar axis, the simulation run at constant $CFL=1$ yielded
a time averaged nondimensional time step $\Delta t \approx 10^{-2}$  thus 
further confirming that the present numerical method alleviates the 
time step restrictions induced by the singularity at the poles.
}

\subsection{Rayleigh--B\'enard convection with central gravity}
\label{sec:RBC}

In this numerical example we use the same configuration as in the previous section 
except for the gravity that now points towards the centre of the sphere. 

\begin{figure}[htb!]
\begin{center} 
\includegraphics[width= 0.98 \textwidth]{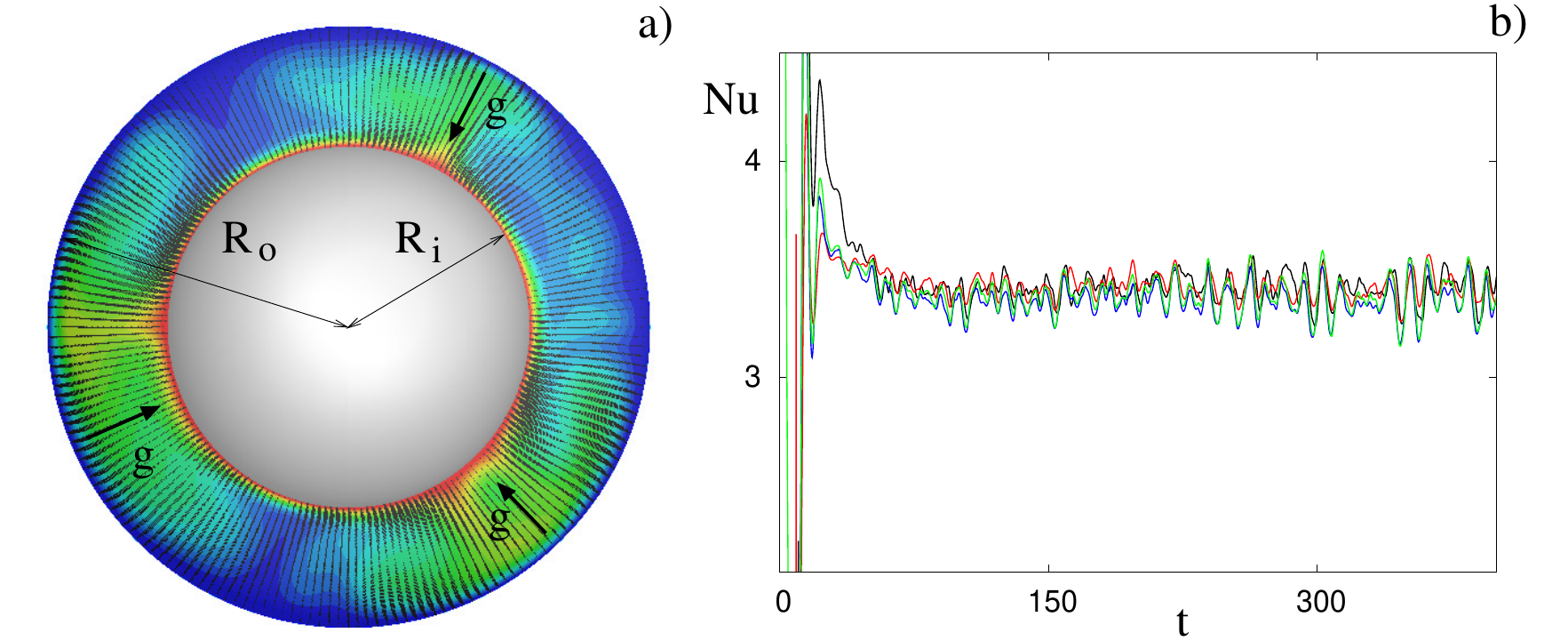}
\end{center}
\caption{a) Section through the meridional planes $\theta=0,\pi$ of temperature 
         contours ($\Delta T = 0.1$) overlaid with velocity vectors 
         for the flow at $\eta=R_i/R_o=0.6$,
$Ra=3\times 10^4$ and $Pr=1.$. Grid $129\times 97\times 97$.
         b) Time evolution of the Nusselt numbers: \solid value computed 
at the inner surface, {\color{red} \solid } value computed at the outer surface,
{\color{green} \solid } value computed from $\epsilon_u$, {\color{blue} \solid } value computed from $\epsilon_T$.
}
\label{fig:RB2}
\end{figure}

We rely again on Equations (\ref{eq:Bou}) with the non-dimensional gravity vector that now
reads ${\bf g}=
(f_\theta, f_r, f_\phi) = (0, g'(r), 0)$ with $g'(r)=(R_o/r)^2$ the non-dimensional radial 
distribution of gravity. The reason for choosing this specific radial dependence is that,
as shown by \cite{Gasti15}, in this case it is possible to derive exact relations
among the Nusselt number and the dissipations:
\begin{equation}
Nu = \frac{Pr^2}{Ra}\frac{1+\eta+\eta^2}{3} \epsilon_u + 1 = 
\frac{1+\eta+\eta^2}{3\eta} \epsilon_T,
\label{eq:diss}
\end{equation}
with $\epsilon_u = <(\nabla \times {\bf u})^2>$ the kinetic energy-- 
and $\epsilon_T = <(\nabla T)^2>$ the temperature variance--dissipation rates
averaged over the fluid 
volume and in time, that can be used to verify both,
the consistency of the numerical method and to assess the statistical convergence
of the results.
In Fig. \ref{fig:RB2} we report the results for the case at $\eta=R_i/R_o=0.6$,
$Ra=3\times 10^4$ and $Pr=1.$ computed on a $129\times 97\times 97$ mesh. It can
be observed that, after the initial transient ($t\leq 100$) the Nusselt numbers
computed from Equations (\ref{eq:Nu}) and (\ref{eq:diss}) oscillate around a common
mean value and the simulation is stopped when their averages agree within $1\%$.

In addition to the heat transfer also the strength of the flow is used to quantify
the response of the system and, in non-dimensional form, it can be expressed by the
Reynolds number. In \cite{Gasti15} it was measured by computing the root mean square
of the velocity field that, however, was scaled by the viscous velocity scale $\nu/(R_o-R_i)$.
Since in Equations (\ref{eq:Bou}) we have used the convective velocity
$\sqrt{g \beta \Delta T (R_o-R_i)}$, the root mean square Reynolds number $Re'$ of
\cite{Gasti15} corresponds to the quantity $\sqrt{2KRa/(VPr)}$ with $V$ the fluid volume
and $K$ the kinetic energy of the flow already defined in section \ref{sec:HV}.

In Table \ref{table:RBG} we report the values of $Nu$ and $Re'$ for some cases that show
excellent agreement with the analogous values obtained by \cite{Gasti15}. In the sake of
conciseness we have not presented the Nusselt numbers obtained by the dissipations from the
expressions (\ref{eq:diss}) that, however, deviate from those computed by the wall temperature
gradients always by less than $1\%$ for every simulation. 

Cases $\#5$--$\#7$ are grid refinements of the same flow: it is worthwhile to note that Case $\#5$ 
has a mesh that is coarser than Case $\#6$ only in the radial direction and the results are still correct.
For the radial distribution of the computational nodes of case $\#5$ we have observed from Equations 
(\ref{eq:Nu}) that being $Nu_i = Nu_o$ the temperature gradient at the inner sphere must exceed that at the
outer sphere by a factor $1/\eta^2$. This implies that the wall resolution at the outer sphere can be
coarser than that at the inner sphere and for the case at $\eta=0.6$, $Pr=1$ and $Ra=3.0\times 10^4$ 
this resulted in a saving of about $25\%$ of nodes. 
In this case, in particular, the radial distribution of the nodes has been assigned as an 
input from an external file built by third--order splines with the conditions $\Delta r =
(R_o-R_i)/450$ at the inner sphere, $\Delta r = (R_o-R_i)/200$ at the outer sphere and
$\Delta r = (R_o-R_i)/45$ halfway between the boundaries.

For these simulations, run only for validation
purposes, this is not a crucial advantage since they can be run anyway within a few hours on a single
Intel I7--2.7GHz processor. However, when the method is employed to tackle higher Rayleigh number
flows implying meshes with hundreds of million (\cite{Gasti15}) or billions (\cite{Steve10}) of nodes
the asymmetric radial stretching of the mesh could become very attractive
 and this is possible thanks to the 
flexibility of the finite--difference schemes.

\begin{table}[ht]
\centering 
\begin{tabular}{c c c c c c c} 
\hline 
Case & $Ra$ & $Nu$ & $Re'$ & $N_\theta \times N_r\times N_\phi$ & $Nu$ \cite{Gasti15}& $Re'$ \cite{Gasti15} \\ [0.5ex] 
\hline\hline 
1 & $1.5\times 10^3$ & $1.327 $ & $4.37$ & $65\times 33\times 49$ & $1.33$ & $4.4$ \\ 
2 & $3.0\times 10^3$ & $1.812 $ & $9.70$ & $65\times 33\times 49$ & $1.80$ & $9.6$ \\ %
3 & $1.0\times 10^4$ & $2.527 $ & $23.43$ & $65\times 33\times 49$ & $2.51$ & $23.3$ \\ %
4 & $1.5\times 10^4$ & $2.828 $ & $29.83$ & $97\times 65\times 65$ & $2.81$ & $29.8$ \\ %
5 & $3.0\times 10^4$ & $3.428 $ & $43.95$ & $97\times 49\times 65$ & $3.40$ & $44.0$ \\ %
6 & $3.0\times 10^4$ & $3.443 $ & $43.94$ & $97\times 65\times 65$ & $3.40$ & $44.0$ \\ %
7 & $3.0\times 10^4$ & $3.412 $ & $43.97$ & $129\times 97\times 97$ & $3.40$ & $44.0$ \\ %
8 & $5.0\times 10^4$ & $3.924 $ & $57.24$ & $97\times 65\times 65$ & $3.89$ & $57.5$ \\ [1ex] 
\hline 
\end{tabular}
\caption{Main input and out parameters and comparison with the results from \cite{Gasti15}.
All the simulations are run at $\eta=0.6$ and $Pr=1$.} 
\label{table:RBG} 
\end{table}

\subsection{Space--developing jet}
\label{sec:SDJ}

\begin{figure}[htb!]
\begin{center} 
\includegraphics[width= 0.98 \textwidth]{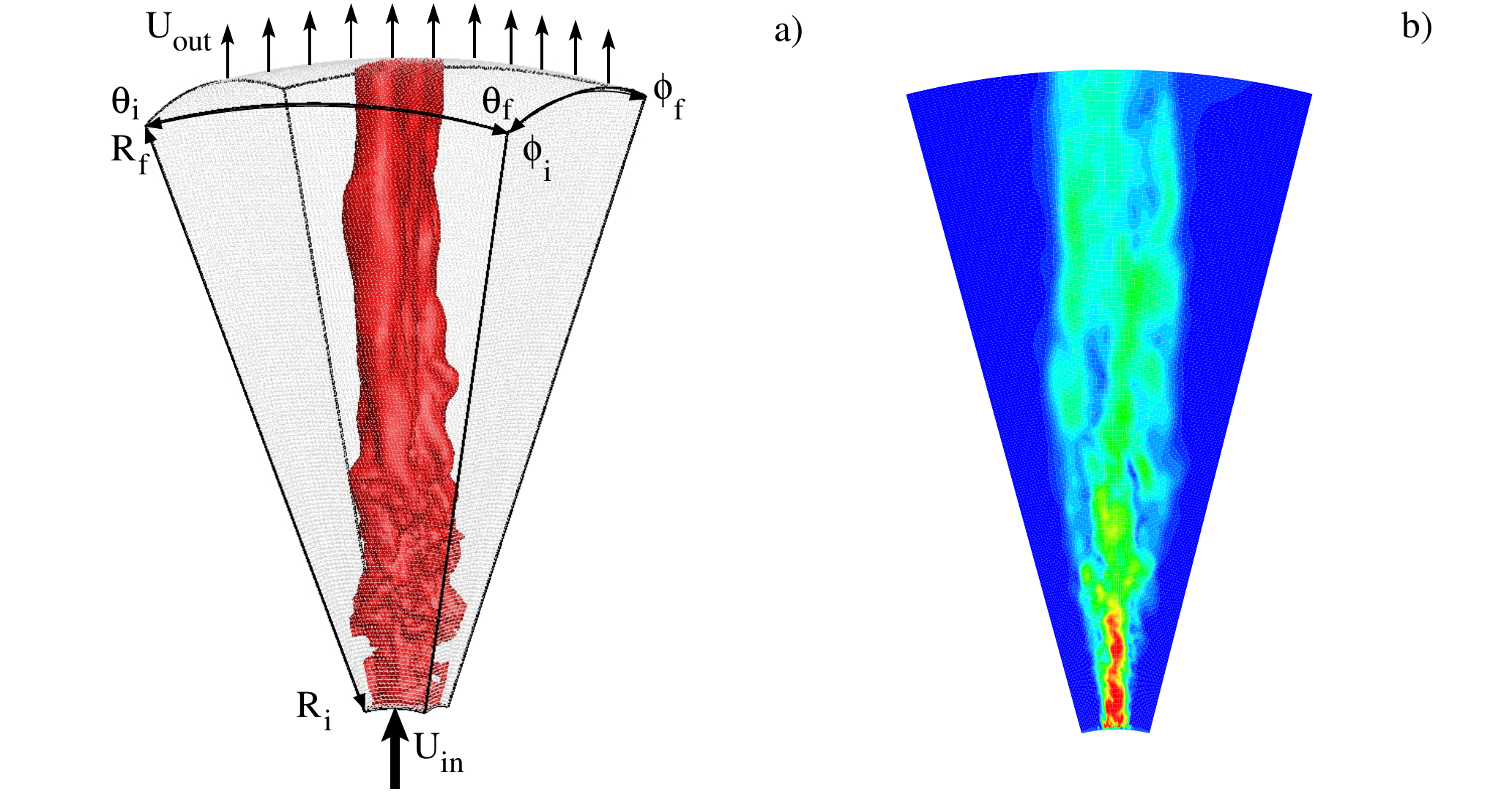}
\end{center}
\caption{a) Computational set--up for the space developing jet. The isosurface is velocity 
magnitude $\mid {\bf u}\mid = 0.2$.
         b) Instantaneous snapshot plane section ($\theta=(\Theta_i+\Theta_f)/2$) of 
velocity magnitude (from blue to red) $\Delta \mid {\bf u}\mid = 0.1$.
}
\label{fig:JET}
\end{figure}

In this last application we simulate the spatial evolution of a round jet of initial diameter $d$ and mean
inflow velocity $U_{in}$ with a Reynolds number $Re= U_{in}d/\nu$. 
Here we follow the idea of \cite{Boers98} who noted that the coordinate lines
of a spherical shell sector naturally follow the self--similar spreading of a jet. Another advantage is that 
the divergence of the $\theta$-- and $\phi$--isolines yield a more refined mesh for small radii while it coarsens
as the radius increases.
The computational domain, reported in Fig. \ref{fig:JET}a, is defined as 
$0=\Theta_i \leq \theta \leq \Theta_f=\pi/6$, $R_i = 3 \leq r \leq 12= R_o$,
and $11\pi/12 = \Phi_i \leq \phi \leq \Phi_f=13\pi/12$ and, since it does not contain any of the singular
points, the solution of the governing equations is performed easily.
In fact, the reason for performing this last test case has not to do with the equation 
singularity but rather with the possibility of the scheme to deal with `complex' boundary
conditions.

At the inner boundary ($r=R_i$) a radial velocity profile with mean $U_{in}$ 
is prescribed within the circle of diameter
$d$ and centre ($[\Theta_i+\Theta_f]/2, [\Phi_i+\Phi_f]/2$) and perturbed with a white noise of amplitude
$0.02U_{in}$. At the outflow we impose the convective boundary condition as in \cite{Salve95}:
\begin{equation}
\frac{\partial q_i}{\partial t} + U_R \frac{\partial q_i}{\partial r} = 0,
\label{eq:bcco}
\end{equation}
that advects all the velocity components $q_i$ out of the domain with the velocity $U_R$ that is 
dynamically adjusted to assure mass conservation to the machine precision.
Periodicity is imposed in the azimuthal and colatitude directions.

The simulation has been performed on a mesh of $97\times 257\times 97$ nodes at a Reynolds number
$Re= 5000$ and an instantaneous snapshot of the velocity magnitude through the mean plane
$\theta=(\Theta_i+\Theta_f)/2$ is shown in Fig. \ref{fig:JET}b.

We wish to point out that our azimuthal and colatitude periodic boundaries are different 
from those of \cite{Boers98} who used traction--free conditions \cite{Gres91} at these lateral planes.
The latter (obtained by imposing the scalar product between the total stress tensor and the
boundary normal to be zero) are certainly more correct since they allow flow transpiration 
without yielding stress in contrast to periodicity that, representing an infinite
array of jets, unavoidably induces some confinement.
 Owing to this important difference, we do not make any quantitative
claim about the dynamics of the jet and do not attempt comparisons with \cite{Boers98};
we use this calculation only to show the flexibility
of the proposed method that allows within the same ease the imposition of simple, homogeneous
boundary conditions, such as the no--slip of sections \ref{sec:RB} and \ref{sec:RBC},
and the inhomogeneous inflow/outflow conditions of the space--developing jet. 

\section{Conclusions}
\label{sec:co}

In this paper we have shown that
\textcolor{black}{
 the combination of a change of variables with a central--, 
second--order accurate finite--differences on a staggered mesh
and the special treatment of some discrete terms
}
removes the singularities of the Navier--Stokes equations for an incompressible viscous flow in 
spherical coordinates. 

Some numerical applications have been considered with the aim of stressing the stability of 
the scheme and its capability to reproduce reference results.

Most of the tests have been performed using a spherical Hill vortex that, in the inviscid limit,
is an exact solution of the governing equations and propagates along a rectilinear trajectory with
a constant velocity. The numerical procedure has shown to be second--order accurate and
to reproduce the results obtained for the same flows by a code in Cartesian coordinates.
The method performed equally well even when the vortex centre was offset with respect to the symmetry
plane $\phi = \pi/2$ or the mesh was unnecessarily refined around the singular points.

\textcolor{black}{
Another benchmark has been performed by simulating the flow within a precessing,
spinning sphere for the same parameters as those considered by
\cite{Kida08} and \cite{Auti09} obtaining a perfect agreement with their
results produced by spectral methods.
}

Equally good results have been obtained for thermally driven flows in which only the singularity at the
polar axis was present although the flow physics was enriched by the presence of the additional temperature 
field; the heat transfer of these flows was always in excellent agreement with other similar studies
and even the exact relations between heat transfer and dissipation rates were perfectly satisfied.

\textcolor{black}{
An important drawback related to the spherical coordinates
is the time step limitation introduced by the discretization around
the singularities. The proposed numerical method has shown to 
alleviate this problem for the polar axis and, for the flow in 
spherical shells with the grid stretched radially at the solid
boundaries, the restriction induced by the latter outweighs that 
of the former.
On the other hand, the singularity at the sphere centre still
introduces strong time step limitations, although only if the largest
flow velocity occurs there.
}

Although the merits of the proposed numerical method have been evidenced by reproducing
simple canonical flows, for which benchmark results are available,
its main merits are related to the flexibility of finite--differences.
In the evolution of a space developing round jet (Section \ref{sec:SDJ}) 
we have qualitatively shown 
the possibility to use complex inflow/outflow 
boundary conditions while for the thermal convection with central gravity we have
employed generic nonuniform meshes (Case \# 5 of Section \ref{sec:RBC}).

These finite--difference features are particularly appealing if the code has to be 
applied to realistic flows,
such as the mantle convection of a planet \cite{Schub04}, in which complex boundary
conditions, but also inhomogeneous forcings and variable fluid properties,
 have to be accounted for.

\textcolor{black}{
Another important advantage of finite--difference methods is the relative ease
of parallelization related to the local nature of the discrete differencing;
this is true also for the present scheme inspired by that of 
\cite{Verzi96} and with 
the same variable arrangement and memory layout.  The latter has been
massively parallelized in \cite{Steve10} and run on up to $3.2\times 10^4$
cores. Also the present scheme in spherical coordinates has been parallelized
and it is already running on hundreds of processors to simulate thermally
driven turbulent flows; these results will be the subject of a
forthcoming paper.
}




\bibliographystyle{model1-num-names}
\bibliography{<your-bib-database>}



\end{document}